\documentclass[apj]{emulateapj}
\usepackage{apjfonts}
\usepackage{natbib}
\usepackage{url}
\usepackage{amsmath}

\input{colordvi.tex}
\usepackage{color}

\slugcomment{ApJ, submitted}

\shorttitle{CRs and non-thermal emission due to cold accretion}
\shortauthors{Inoue, Uchiyama, Arakawa, Renaud \& Wada}

\begin{document}

\title{Cosmic Rays and Non-thermal Emission Induced by Accretion of Cool Gas onto the Galactic Disk}

\author{Susumu Inoue\altaffilmark{1}, Yasunobu Uchiyama\altaffilmark{2}, Masanori Arakawa\altaffilmark{2,1}, Matthieu Renaud\altaffilmark{3}, and Keiichi Wada\altaffilmark{4}}

\altaffiltext{1}{Astrophysical Big Bang Laboratory, RIKEN, 2-1 Hirosawa, Wako 351-0198 Japan; susumu.inoue@riken.jp}
\altaffiltext{2}{Department of Physics, Rikkyo University, 3-34-1 Nishi-Ikebukuro, Toshima-ku, Tokyo 171-8501, Japan}
\altaffiltext{3}{Laboratoire Univers et Particules de Montpellier, Universit\'e Montpellier, CNRS/IN2P3, CC 72, Place Eug\`ene Bataillon, F-34095 Montpellier Cedex 5, France}
\altaffiltext{4}{Graduate School of Science and Engineering, 1-21-35 Korimoto, Kagoshima University, Kagoshima 890-0065, Japan}

\begin{abstract}
On both observational and theoretical grounds, the disk of our Galaxy
should be accreting cool gas with temperature $\la 10^5$ K via the halo at a rate $\sim 1 \ {\rm M_\sun \ yr}^{-1}$.
At least some of this accretion is mediated by high velocity clouds (HVCs),
observed to be traveling in the halo with velocities of a few 100 km s$^{-1}$
and occasionally impacting the disk at such velocities, especially in the outer regions of the Galaxy.
We address the possibility of particle acceleration in shocks triggered by such HVC accretion events,
and the detectability of consequent non-thermal emission in the radio to gamma-ray bands and high-energy neutrinos.
For plausible shock velocities $\sim {\rm 300 \ km \ s^{-1}}$ and magnetic field strengths $\sim 0.3 - 10 \ {\rm \mu G}$,
electrons and protons may be accelerated up to $\sim 1-10$ TeV and  $\sim 30 - 10^3$ TeV, respectively,
in sufficiently strong adiabatic shocks during their lifetime of $\sim 10^6 \ {\rm yr}$.
The resultant pion decay and inverse Compton gamma-rays may be the origin
of some unidentified Galactic GeV-TeV sources,
particularly the ``dark'' source HESS J1503-582 that is spatially coincident with anomalous HI structure known as ``forbidden-velocity wings''.
Correlation of their locations with star-forming regions may be weak, absent, or even opposite.
Non-thermal radio and X-ray emission from primary and/or secondary electrons
may be detectable with deeper observations.
The contribution of HVC accretion to Galactic cosmic rays is subdominant,
but could be non-negligible in the outer Galaxy.
As the thermal emission induced by HVC accretion is likely difficult to detect,
observations of such phenomena may offer a unique perspective on
probing gas accretion onto the Milky Way and other galaxies.
\end{abstract}


\section{Introduction}
\label{sec:intro}

\subsection{Gas accretion onto galaxies}
\label{sec:gasacc}

Gas accretion onto galaxies over cosmological timescales
is understood to be one of the key processes governing their formation and evolution,
as deduced from a variety of observational and theoretical considerations
(for reviews, see e.g. \citealt{Sancisi08, Putman12, Sanchez14, Fox17}).
To sustain the star formation rate of the Milky Way
that is inferred to have been approximately constant over the last several Gyr,
the Galactic disk must have been continuously supplied with gas at a rate of order $\dot{M}_{\rm acc, SF} \sim 1 \ {\rm M_\sun \ yr}^{-1}$.
Potential sources of the accreting gas include the intergalactic medium (IGM) channeled along filaments of the cosmic web,
gas stripped from satellite galaxies, condensation out of hot halo gas via thermal instabilities,
and gas recycled from supernova (SN)-driven fountains or winds from the disk \citep{Oort70, Shapiro76, Richter17}.
The observed metallicity distribution of long-lived disk stars
indicates that such accreting gas must be relatively metal-poor \citep{Larson72, Finlator17},
favoring the IGM, satellite and/or halo gas as the dominant source.
This is supported by cosmological simulations \citep[e.g.][]{Fernandez12, Joung12warm, Nuza14},
although SN-driven gas may still play an important auxiliary role \citep{Fraternali17}.
Similar inferences can be made for other actively star-forming galaxies at all redshifts (\citealt{Putman17, Lockman17} and references therein).

Theoretically, the classical picture of galaxy formation dictated that
most of the collapsing gas is initially shock heated to the virial temperature of the gravitational potential
(\citealt{Rees77, Silk77, White78});
a view that continued after the emergence of the cold dark matter (CDM) cosmology \citep{Blumenthal84, White91}.
However, more recent studies led to the recognition that
substantial infall of cold gas can occur via filamentary streams without being shock heated until it reaches
the denser regions of the forming galaxy (\citealt{Birnboim03, Keres05, Stewart17}; see also \citealt {Binney77}).
Such ``cold accretion'' may be the dominant mode of gas accretion
for all galaxies at early epochs, as well as in present-day galaxies with total mass $M \la 10^{12} {\rm M_\sun}$,
close to that of the Milky Way (e.g. \citealt{Brooks09, vdVoort11, Faucher11}; see however \citealt{Nelson13}).

\subsection{High velocity clouds and accretion}
\label{sec:HVCacc}

The most direct evidence of infalling cold gas (with temperature $T \la 10^4 {\rm \ K}$) in the Milky Way is provided by
high-velocity clouds (HVCs) of neutral hydrogen, observed with line-of-sight (LOS) velocities
between -500 and +450 ${\rm  km \ s^{-1}}$ in the local-standard-of-rest frame
that deviate strongly from simple expectations for Galactic rotation
\citep{Wakker97, Putman12, Richter17}.
They are found throughout the sky with a range of angular sizes,
from $< 2$ deg for compact HVCs (CHVCs)
up to $\sim 1500 {\rm \ deg}^2$ for large HVC complexes,
a range of HI column densities $N_{\rm HI} \ga 10^{17}-10^{20} {\rm \ cm^{-2}}$,
metallicities (relative to solar) $Z/Z_\sun \sim 0.1-0.5$, and lacking stellar components.
While a major fraction of HVCs is known to constitute the Magellanic Stream and the Leading Arm (hereafter collectively MS),
likely comprising gas stripped from the Magellanic Clouds \citep{DOnghia16},
the origin of most other HVCs is unclear.
Many of the larger complexes have been constrained to reside in the halo,
with distances $D \sim 2-15 {\rm \ kpc}$ and HI masses $M_{\rm HI} \sim 10^5 - 5 \times 10^6 \ {\rm M_\sun}$,
implying mean densities $n_{\rm HI} \sim 0.1 \ {\rm cm^{-3}}$ \citep{Putman12, Richter17}.
Most HVCs are also known to possess associated, warm ionized gas (with $T \sim 10^4 - 10^5 {\rm \ K}$)
observed in H$\alpha$ or low ionization metal lines,
whose total mass may be greater than that in HI \citep{Lehner11, Putman12, Richter17}.
HVCs have also been identified in nearby galaxies including M31 and M33 \citep[][and references therein]{Lockman17}.

Long suspected to be the fuel reservoir for disk star formation \citep{Oort70},
the net gas accretion rate due to HVCs is uncertain,
and recent estimates vary among different authors.
Depending on the treatment of the ionized component and the contribution from the MS,
it ranges from $\dot{M}_{\rm acc, HVC} \sim 0.1\ {\rm M_\sun \ yr^{-1}}$ \citep{Putman12}
to $\dot{M}_{\rm acc, HVC} \ga 5 \ {\rm M_\sun \ yr^{-1}}$ \citep{Richter17},
between falling significantly short to being more than sufficient for the required $\dot{M}_{\rm acc, SF}$.
The dominance of warm ionized gas over HI for accretion
appears to be supported by recent cosmological simulations \citep{Joung12warm, Murante12, Nuza14}.
Hydrodynamical simulations suggest
that HVCs entering the inner regions of the halo may be substantially disrupted and ablated by interacting with ambient hot gas
(at $T \ga 10^6 {\rm \ K}$) that likely permeates the halo \citep{Spitzer56, BlandHawthorn16},
unless their initial gas mass $M \ga 10^{4.5} \ {\rm M_\sun}$
\citep{Heitsch09, Kwak11, Joung12cloud, Armillotta17}.
SN-driven outflows may further disturb HVCs near the disk,
although mixing with metal-enriched gas from the former may allow remnant gas from the latter
to eventually recool and undergo ``quiet'' accretion at low velocities \citep{Marinacci10, Fraternali17}.

Direct accretion of ``cool'' gas with $T \la 10^5 {\rm \ K}$ (here referring to both cold and warm gas)
onto the disk at high velocities is expected to be most important in the outer Galaxy,
where HVCs are likely less affected by the hot halo
or outflows from the disk driven by SNe or the central supermassive black hole (SMBH; \citealt{vdVoort17}).
Moreover, gas from all likely sources (IGM, satellites or halo) may be more prone to accrete
onto the outer disk regions due to their higher angular momenta relative to the disk \citep{Peek09, Christensen16, Stewart17},
which may be subsequently transported radially inward to fuel star formation in the central regions \citep{Finlator17}.
This notion may be consistent with various observations \citep{Combes14}
and cosmological simulations \citep{Stewart11, Fernandez12}.
Direct collisions of HVCs with the disk have been proposed
as the origin of supershells and other large HI structures in the disk
that cannot be easily explained by stellar activity or supernovae \citep{Heiles84, Tenorio88, Baek08},
as well as warps and lopsidedness seen in the outer regions of most disk galaxies including the Milky Way \citep{Sancisi08, Combes14}.

A particularly clear example of a direct HVC accretion event was recently discovered by \cite{Park16}.
It was initially identified as a faint ``forbidden-velocity wing'' (FVW), i.e. localized HI structure in the Galactic disk
with velocity deviating from Galactic rotation by more than $~20 {\rm \ km \ s^{-1}}$ \citep{Kang07}.
Deeper HI observations at higher spatial and spectral resolution revealed a kpc-scale supershell
with a CHVC at its center,
the former most likely resulting from the impact of the latter with the disk.
\cite{Park16} favor a location in the outer Galaxy at Galactocentric radius $R \sim 15 {\rm \ kpc}$,
and infer initial values for the relative velocity $v \sim 240 {\rm \ km \ s^{-1}}$
and HVC mass $M_{\rm HVC} \ga 6 \times 10^4 \ {\rm M_\sun}$.

\subsection{HVC accretion events and non-thermal phenomena}
\label{sec:HVCnt}

HVCs striking the disk without much prior deceleration
will give rise to forward shocks in the disk gas and/or reverse shocks within the HVC
that can be strong depending on the conditions
\citep[e.g.][]{Tenorio86, Tenorio87, Baek08}.
Although the thermal emission from the shocked gas expected in the UV to soft X-ray band
is likely difficult to observe,
an interesting question is whether such shocks can generate
high-energy protons and electrons via diffusive shock acceleration (DSA; e.g. \citealt{Bell78, Blandford78})
and induce non-thermal emission that may be observable.
Of particular interest in this regard are numerous unidentified GeV-TeV gamma-ray sources,
many spatially extended,
that have been discovered in surveys of the Galactic Plane by
H.E.S.S. \citep{Aharonian05, Aharonian06},
{\it Fermi}-LAT \citep{Ackermann16, Ackermann17_ES}, HAWC \citep{Abeysekara17}, and other facilities.
Although a fair fraction of TeV sources from the first surveys were later revealed to be known types of objects
such as supernova remnants (SNRs; \citealt{Gottschall16}) or pulsar wind nebulae (PWNe; \citealt{Abdalla17}),
some have remained mysterious,
and new unidentified sources continue to be found with deeper observations
(\citealt{Deil15, Donath17}; see TeVCat \footnote{\url{http://tevcat.uchicago.edu/}}).
One especially enigmatic source is HESS J1503-582,
detected up to $\sim$ 20 TeV with spatial extension $\sim 0.5 ^\circ$ \citep{Renaud08}
and a sub-TeV counterpart \citep{Ackermann16}.
Categorized as a ``dark'' source in TeVCat
with no obvious counterparts at radio, infrared or X-ray bands,
\cite{Renaud08} report a possible association with a FVW of unknown distance and nature,
initially discovered by \cite{Kang07}.

Here we discuss the possibility that shocks driven by HVC accretion events of the kind described above
are accelerators of protons and electrons to GeV-TeV energies
and consequent sources of non-thermal emission in gamma-rays and other wavebands.
Analogous to the FVW that was revealed to be a CHVC-supershell system \citep{Park16},
we propose that HESS J1503-582 with its associated FVW was triggered by a similar event in the outer Galactic disk,
its gamma-rays resulting from neutral pion-decay and inverse Compton emission induced by protons and electrons, respectively.
We note that HVCs at high Galactic latitudes not interacting with the disk are undetected in gamma rays,
a fact has has been used to constrain the amount of cosmic rays (CRs) propagating in the halo
(\citealt{Tibaldo15}; see however \citealt{Blom97}).

The rest of the paper is organized as follows.
The next three sections describe
basic aspects of physics relevant to HVC accretion events,
our assumptions and parameters, and some order-of-magnitude estimates,
concerning shock properties in \S \ref{sec:shock},
particle acceleration in \S \ref{sec:accel},
and non-thermal emission in \S \ref{sec:emission}.
In \S \ref{sec:HVCgam}, the expected non-thermal emission is calculated numerically
and compared with observations of the ``dark'' GeV-TeV source HESS J1503-582.
Further observational tests and searches for other similar sources with current and future facilities
are discussed in \S \ref{sec:tests}.
We conclude in \S \ref{sec:conc} 
by touching upon implications for
the observed diffuse gamma-ray emission in the outer Galaxy,
and prospects for obtaining new insight into gas accretion processes onto galaxies.
Most salient points of this work were already presented
at the 5th International Symposium on Gamma-Ray Astronomy in 2012, for which the slides are available online
\footnote{\url{https://www.mpi-hd.mpg.de/hd2012/pages/presentations/Inoue.pdf}}.

\section{Shock properties and energetics}
\label{sec:shock}

\subsection{HVCs and the Galactic disk}
\label{sec:HVCdisk}

Key parameters for HVCs are its 3-dimensional velocity
that can be characterized by magnitude $v_{\rm HVC}$ and orientation relative to the disk rotation,
total gas mass $M_{\rm HVC}$ including both cold and warm ionized components at $T \la 10^5 {\rm \ K}$,
and mean internal gas density $n_{\rm HVC}$,
just before their impact with the disk.
One also needs the distribution of all these variables
and their spatial dependence, particularly on Galactocentric radius $R$, for the population of HVCs.
For HI gas in the disk, plausible models are available
for the mean radial profiles of circular velocity $v_{\rm disk}$, density $n_{\rm disk}$ and scale height $h_{\rm disk}$ 
that are in accord with existing observations.
The rotation curve is consistent with being roughly flat at $v_{\rm disk} \sim 240 {\rm \ km \ s^{-1}}$
from the solar circle at $R = R_\sun \simeq 8.3$ kpc out to at least $R \sim 16$ kpc \citep{Reid14},
albeit with large observational uncertainties \citep{BlandHawthorn16}.
The HI disk is known to extend significantly beyond the observed stellar disk,
possibly up to $R \sim 60$ kpc \citep{Kalberla09}.
The warm ionized gas in the disk may also be relevant, but is not considered here
as its surface density is subdominant, notably in the outer Galaxy \citep{Ferriere01, Yao17}.
With knowledge of these quantities,
we can estimate for each collision the relative velocity $v_{\rm acc}$ between the HVC and the disk,
and the mean velocities $v_{s,f}$ and $v_{s,r}$
for the resultant forward shock (FS) in the disk and the reverse shock (RS) in the HVC, respectively.
If the cloud's initial gas mass $M \ga 10^{4.5} \ {\rm M_\sun}$ and/or the location is in the outer Galaxy,
HVCs may not be significantly decelerated before disk impact (\S \ref{sec:HVCacc}),
in which case $v_{\rm HVC}$ is expected to be of order $v_{\rm disk}$,
especially for HVCs originating from the IGM or satellites high above the Galactic Plane.
Consequently, $v_{\rm acc}$ should also be of the same order.

The actual distribution of the HVC parameters is highly uncertain.
Observations only provide the LOS component of the velocity, and are strongly affected by selection effects near the disk. 
Measurements of $M_{\rm HVC}$ and $n_{\rm HVC}$ are only available for the largest HVCs with distance constraints.
Current cosmological simulations do not yet have sufficient resolution to elucidate the detailed physics of the disk-halo interface.
In view of the uncertainties, we simply regard as parameters $v_{\rm HVC}$ and $v_{\rm acc}$
with the same fiducial value of $300 {\rm \ km \ s^{-1}}$.
We also fiducially take $n_{\rm HVC} = 0.1 \ {\rm cm^{-3}}$ and $M_{\rm HVC} = 10^5 \ {\rm M_\sun}$,
the latter above the inferred threshold for HVC survival from disruptive hydrodynamical processes in the halo.
Finally, we consider a fiducial impact location at $R = 15 \ {\rm kpc}$,
where $n_{\rm disk} \sim 0.1 \ {\rm cm^{-3}}$ at midplane and $h_{\rm disk} \sim 300 \ {\rm pc}$ for the HI disk \citep{Kalberla09}.
Our fiducial values are in line with those invoked for the pre-collision CHVC by \cite{Park16}.

\subsection{Shock velocity and lifetime}
\label{sec:shockvt}

Under the idealized assumption of uniform spherical clouds colliding perpendicularly with a static, uniform slab disk
with given $v_{\rm acc}$, $n_{\rm HVC}$ and $n_{\rm disk}$,
approximate analytic descriptions of the relevant hydrodynamics have been presented by \cite{Tenorio80, Tenorio81}.
In reality, this only applies in the rare case when the HVC has no radial motion and its azimuthal velocity equals $v_{\rm disk}$,
not to mention further complications in situations with oblique incidence, non-trivial density structure and geometry, etc.
Nevertheless, we can expect that both $v_{s,f}$ and $v_{s,r}$ will be of order $v_{\rm acc}$,
as long as $n_{\rm HVC}/n_{\rm disk}$ is not greatly different from unity.
We consider $v_s$ as a parameter to stand for either $v_{s,f}$ or $v_{s,r}$, again with fiducial value $300 {\rm \ km \ s^{-1}}$,
bearing in mind that for $v_s \la 120 {\rm \ km \ s^{-1}}$,
particle acceleration to high energies may be suppressed by non-trivial effects (\S \ref{sec:neutral}).
A more detailed description warrants hydrodynamical simulations \citep[e.g.][]{Tenorio86, Tenorio87, Kudoh04, Baek08},
which are beyond the scope of this paper.

The sonic Mach number of a shock propagating into gas with temperature $T_0$ is
\begin{multline}
  {\cal M}_s = {v_s \over (\gamma k_B T_0 / \mu m_p)^{1/2}} \\
                   \simeq 20 \left(\mu \over 0.61\right)^{1/2} \left(v_s \over 300 \ {\rm km \ s^{-1}}\right) \left(T_0 \over 10^4 \ {\rm K}\right)^{-1/2} ,
  \label{eq:Mach_s}
\end{multline}
where $k_B$ is the Boltzmann constant, $m_p$ is the proton mass,
$\gamma=5/3$ is the adiabatic index for monatomic gas, and $\mu$ is the mean molecular weight.
The latter is respectively $\mu=0.61$ or $\mu=1.28$ for fully ionized or neutral gas with $Z/Z_\sun \simeq 0.2$,
the typical metallicity for our conditions (see below).
The disk gas at $R > R_\sun$ is suspected to be mainly in the warm neutral phase with $T_0 \sim 6000-10^4 \ {\rm K}$ \citep{Ferriere01},
so that for HVC-driven FS in this region,
${\cal M}_s \sim 30-40 \ (v_s/300 \ {\rm km \ s^{-1}})$,
or higher if the disk gas at the impact location is colder.
On the other hand, observed HI line widths of HVCs indicate $T_0 \sim 500-9000 \ {\rm K}$ \citep{Putman12}, so
${\cal M}_s \sim 30-130 \ (v_s/300 \ {\rm km \ s^{-1}})$
for RS within the cold component of HVCs, while
${\cal M}_s \sim 6-20 \ (v_s/300 \ {\rm km \ s^{-1}})$
for those in the warm ionized component with $T_0 \sim 10^4-10^5 {\rm K}$.
Most of this range qualify as strong shocks with compression ratio $r_c=(\gamma+1) {\cal M}_s^2 / [(\gamma-1) {\cal M}_s^2 +2] \sim 4$,
and can be interesting for particle acceleration (\S \ref{sec:acceleff}).

The radius of a uniform, spherical HVC is
\begin{multline}
  r_{\rm HVC} = \left(3 M_{\rm HVC} \over 4 \pi m_p n_{\rm HVC}\right)^{1/3} \\
                   \simeq 213 \ {\rm pc} \ \left(n_{\rm HVC} \over 0.1 \ {\rm cm^{-3}}\right)^{-1/3} \left(M_{\rm HVC} \over 10^5 \ {\rm M_\sun}\right)^{1/3} .
  \label{eq:r_HVC}
\end{multline}
Assuming that the shocks remain adiabatic,
the approximate crossing time of the RS through the HVC is
\begin{multline}
  \tau_{\rm HVC} \approx {2 r_{\rm HVC} \over v_{s,r}}
                           \simeq 1.4 \times 10^6 \ {\rm yr} \\
                           \times \left({n_{\rm HVC} \over 0.1 \ {\rm cm^{-3}}}\right)^{-1/3} \left({M_{\rm HVC} \over 10^5 \ {\rm M_\sun}}\right)^{1/3}
                                      \left(v_s \over 300 \ {\rm km \ s^{-1}}\right)^{-1} ,
  \label{eq:t_HVC}
\end{multline}
while that for the FS through the disk is
\begin{multline}
  \tau_{\rm disk} \approx {2 h_{\rm disk} \over v_{s,f} \cos \theta}
                          \simeq 2.0 \times 10^6 \ {\rm yr} \\
                          \times {1 \over \cos \theta} \left({h_{\rm disk} \over 300 {\rm \ pc}}\right) \left(v_s \over 300 \ {\rm km \ s^{-1}}\right)^{-1} ,
  \label{eq:t_disk}
\end{multline}
where $\theta$ is the angle of impact with respect to the disk normal.

The post-shock gas temperature for a strong shock is
\begin{equation}
  T_s = {2 (\gamma -1) \mu v_s^2 \over (\gamma +1)^2 k_B}
         \simeq 1.3 \times 10^6 \ {\rm K} \left(v_s \over 300 \ {\rm km \ s^{-1}}\right)^2 .
  \label{eq:T_s}
\end{equation}
The thermal emission peaking in the far UV to soft X-ray range
could be challenging to observe due to strong photoelectric absorption by the foreground interstellar medium (ISM; \S \ref{sec:thermal}).
For gas in collisional ionization equilibrium at constant density $n$, $10^5 < T <10^{6.5} \ {\rm K}$ and $0.1 < Z/Z_\sun < 1$,
the radiative cooling time can be approximated by
$\tau_{g, {\rm rad}} \sim 1.3 \times 10^{5} \ {\rm yr} \ (n/ 1 \ {\rm cm^{-3}})^{-1} \ (Z/Z_\sun)^{-0.8} (T/ 10^6 \ {\rm K})^{1.7}$ \citep{Draine11}.
The radiative cooling time of the post-shock gas is
\begin{multline}
  \tau_{g, {\rm rad}} \simeq 1.8 \times 10^6 \ {\rm yr} \\
            \times \left({r_c n_0 \over 0.4 \ {\rm cm^{-3}}}\right)^{-1} \left(Z \over 0.2 \ Z_\sun\right)^{-0.8} \left(v_s \over 300 \ {\rm km \ s^{-1}}\right)^{3.4} ,
  \label{eq:t_grad}
\end{multline}
where $n_0$ denotes the pre-shock gas density.
Note that $Z/Z_\sun \sim 0.1-0.3$ is the typical observed metallicity of HVC gas \citep{Putman12, Richter17},
while $Z/Z_\sun \sim 0.2-0.3$ for the disk at $R \sim 15 \ {\rm kpc}$ (\citealt{Henry99, Matteucci14} and references therein).

The duration of the phase during which the shocks traverse the medium at roughly constant velocity before beginning to decelerate
can be estimated by
$\tau_s \sim \min[\tau_{\rm disk}, \tau_{\rm HVC}, \tau_{g, {\rm rad}}]$.
If $\tau_{\rm HVC} < \tau_{\rm disk} < \tau_{g, {\rm rad}}$,
the entire HVC is shocked and decelerated by the adiabatic RS before the FS crosses the disk,
after which the RS decays and the FS decelerates.
If $\tau_{\rm disk} < \tau_{\rm HVC} < \tau_{g, {\rm rad}}$,
the adiabatic FS driven by the HVC emerges from the opposite side of the disk before the cloud is completely shocked,
after which the FS decays and the RS decelerates.
When $\tau_{g, {\rm rad}}$ is shortest timescale, both FS and RS become radiative and start to decelerate
before crossing the disk and cloud, respectively.
Note that for given $v_{\rm acc}$,
$v_{s,f}$ and $v_{s,r}$ differ by a numerical factor that depends on $n_{\rm HVC}/n_{\rm disk}$ \citep{Tenorio80, Tenorio81},
so strictly speaking, $\tau_{\rm HVC}$ and $\tau_{\rm disk}$ should not be evaluated simultaneously with the same value of $v_s$.
This pre-deceleration phase at constant velocity will be our main focus,
even though the ensuing deceleration phase may also be of some interest for particle acceleration (\S \ref{sec:neutral}).

\begin{figure}[!htb]
\epsscale{1.2}
\plotone{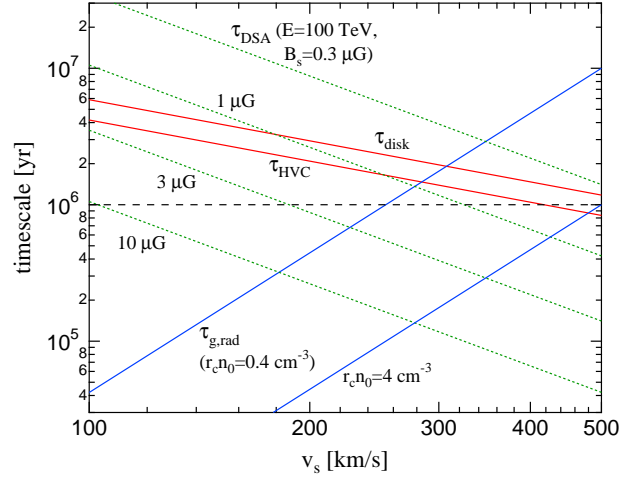}
\caption{
Characteristic timescales versus shock velocity $v_s$.
Plotted are the crossing time of the Galactic disk by the forward shock, $\tau_{\rm disk}$ (Eq. \ref{eq:t_disk}; upper red line),
the crossing time of the high velocity cloud by the reverse shock, $\tau_{\rm HVC}$ (Eq. \ref{eq:t_HVC}; lower red line),
and the radiative cooling time of the post-shock gas, $\tau_{g, {\rm rad}}$, with density $r_c n_0 =$ 0.4 and 4 ${\rm cm^{-3}}$
(Eq. \ref{eq:t_grad}; upper and lower blue lines, respectively).
Overlayed are the timescales for diffusive shock acceleration, $\tau_{\rm DSA}$,
of protons up to energy $E =$ 100 TeV
with post-shock magnetic field $B_s =$ 0.3, 1, 3 and 10 ${\rm \mu G}$
(Eq. \ref{eq:t_accel} with $q=1$; green dashed lines, from top to bottom).
The age of $t = 10^6$ yr considered for our models in \S \ref{sec:HVCgam} is also denoted (black long-dashed line).
}
\label{fig:time}
\end{figure}

The timescales $\tau_{\rm disk}$, $\tau_{\rm HVC}$, and $\tau_{g, {\rm rad}}$
are plotted as functions of $v_s$ in Fig. \ref{fig:time},
showing that unless $v_s \ga 300 \ {\rm km \ s^{-1}}$ for our fiducial case of $r_c n_0 = 0.4 \ {\rm cm^{-3}}$,
both RS and FS become radiative before crossing their respective media,
where particle acceleration is likely less efficient (\S \ref{sec:neutral}).
If $r_c n_0 = 4 \ {\rm cm^{-3}}$, this is unavoidable up to $v_s \sim 500 \ {\rm km \ s^{-1}}$.

\subsection{Energetics and number}
\label{sec:enenum}

The kinetic energy of a HVC is
\begin{multline}
  {\cal E}_{\rm HVC}  =  {1 \over 2} M_{\rm HVC} v_{\rm HVC}^2 \\
        \simeq 9.0 \times 10^{52} \ {\rm erg} \left(M_{\rm HVC} \over 10^5 \ {\rm M_\sun} \right) \left(v_{\rm HVC} \over 300 \ {\rm km \ s^{-1}}\right)^2 ,
  \label{eq:E_HVC}
\end{multline}
which could be up to ${\cal E}_{\rm HVC} \sim 5 \times 10^{54} (v_s / 300 \ {\rm km \ s^{-1}})^2 \ {\rm erg}$ for the most massive HVCs.
Their significantly larger energy compared to typical SN explosions led them
to be recognized as promising energy sources for creating supershells and other large HI structures \citep{Tenorio88}.
The rate at which this kinetic energy flows into a shock with surface area $A_s = \pi r_s^2 \sim \pi r_{\rm HVC}^2$ is
\begin{multline}
  L_{k, {\rm HVC}} \approx {1 \over 2} m_p n_0 v_s^3 A_s
  \simeq 3.1 \times 10^{39} \ {\rm erg \ s^{-1}} \\
  \times  \left(n_0 \over 0.1 \ {\rm cm^{-3}}\right) \left(n_{\rm HVC} \over 0.1 \ {\rm cm^{-3}}\right)^{-2/3} \left(M_{\rm HVC} \over 10^5 M_\sun \right)^{2/3} \left(v_s \over 300 \ {\rm km s^{-1}}\right)^3 ,
  \label{eq:L_kin}
\end{multline}
where $n_0 = n_{\rm HVC}$ for the RS and $n_0 = n_{\rm disk}$ for the FS, and Eq. \ref{eq:r_HVC} has been used.
For $n_0 = n_{\rm HVC}$, the expression reduces to
$L_{k, {\rm HVC}} \approx (3/2) ({\cal E}_{\rm HVC}/\tau_{\rm HVC}) (v_s/v_{\rm HVC})^2$.

The total power due to all accreting HVCs in the Galaxy with $v_{\rm acc}$,
i.e. the rate at which their kinetic energy is dissipated in the disk, can be estimated as
\begin{multline}
  {\cal L}_{\rm acc, HVC} \approx  {1 \over 2} f_{\rm acc} \dot{M}_{\rm acc, HVC} v_{\rm acc}^2 \\
                        \simeq 2.9 \times 10^{40} \ {\rm erg \ s^{-1}}
                         f_{\rm acc} \left(\dot{M}_{\rm acc,HVC} \over 1 \ {\rm M_\sun \ yr^{-1}} \right) \left(v_{\rm acc} \over 300 \ {\rm km \ s^{-1}}\right)^2 ,
  \label{eq:L_acc}
\end{multline}
Some or possibly most of this reflects accretion onto the outer regions of the disk (\S \ref{sec:HVCacc}, \S \ref{sec:HIobs}).
The factor $f_{\rm acc}$ accounts for a number of effects.
First, as there must be a distribution in $v_{\rm acc}$,
only a part of $\dot{M}_{\rm acc, HVC}$ corresponds to
direct accretion onto the disk with $v_{\rm acc} \sim 300 \ {\rm km \ s^{-1}}$.
Second, some HVCs may penetrate the disk while dissipating only a fraction of their kinetic energy \citep{Tenorio86, Tenorio87, Baek08}.
A realistic value of $f_{\rm acc}$ is difficult to estimate, but $f_{\rm acc} \ll 1$ is possible
if much of the accretion turns out to proceed quietly with low $v_{\rm acc}$ \citep[e.g.][]{Fraternali17}.
On the other hand, the total power from SNe with kinetic energy ${\cal E}_{\rm SN}$ and rate ${\cal R}_{\rm SN}$ is
\begin{equation}
  {\cal L}_{\rm SN} = {\cal E}_{\rm SN} {\cal R}_{\rm SN}
          \simeq 9.5 \times 10^{41} \ {\rm erg \ s^{-1}} \left({{\cal E}_{\rm SN} \over 10^{51} \ {\rm erg}} {{\cal R}_{\rm SN} \over 0.03 \ {\rm yr^{-1}}}\right) .
  \label{eq:L_SN}
\end{equation}
As is well known, with a plausible CR acceleration efficiency of order 10 \% (\S \ref{sec:acceleff}),
${\cal L}_{\rm SN}$ can account for the energy budget of the observed Galactic CRs
with inferred power ${\cal L}_{\rm CR} \sim 10^{41} \ {\rm erg \ s^{-1}}$ \citep{Ginzburg64, Strong10, Grenier15}.
In comparison, ${\cal L}_{\rm acc, HVC}$ is only $\sim 3$ \% of ${\cal L}_{\rm SN}$,
and possibly even less if $f_{\rm acc} \ll 1$.
Nevertheless, the energetics of HVC accretion can be relatively more important
in the outer Galaxy where ${\cal R}_{\rm SN}$ is much lower (\S \ref{sec:conc}).

The mean number of active HVC accretion events with adiabatic shocks in the Galaxy can be crudely estimated by
\begin{multline}
  N_s  \approx  f_{\rm s} (\dot{M}_{\rm acc, HVC}/M_{\rm HVC}) \tau_s \\
       \simeq 10 \ f_{\rm s} \left({\dot{M}_{\rm acc,HVC} \over 1 \ {\rm M_\sun \ yr^{-1}}} {\tau_s \over 10^6 \ {\rm yr}}\right)
       \left(M_{\rm HVC} \over 10^5 \ {\rm M_\sun}\right)^{-1} ,
  \label{eq:N_s}
\end{multline}
Similar to $f_{\rm acc}$, $f_{\rm s}$ accounts for the fact that only a fraction of $\dot{M}_{\rm acc, HVC}$
is represented by accretion of HVCs with $M_{\rm HVC} \sim 10^5 \ {\rm M_\sun}$ and $\tau_s \sim 10^6 \ {\rm yr}$.

The above estimates of ${\cal L}_{\rm acc, HVC}$ and $N_s$ are subject to uncertainties in $\dot{M}_{\rm acc, HVC}$.
\cite{Putman12} give $\dot{M}_{\rm acc, HVC} \sim 0.1-0.4 \ M_\sun {\rm \ yr^{-1}}$,
including the ionized gas seen in H$\alpha$,
but excluding the contribution from the MS.
Estimating the contribution of ionized gas detected via metal lines,
\cite{Lehner11} derive $\dot{M}_{\rm acc, HVC} \sim 0.45-1.40 \ M_\sun {\rm \ yr^{-1}}$,
which meets the required $\dot{M}_{\rm acc, SF}$.
Accounting for the MS that may potentially dominate accretion for $\sim$1 Gyr,
\cite{Richter12, Richter17} give $\dot{M}_{\rm acc, HVC} \sim 0.7 \ M_\sun {\rm \ yr^{-1}}$ in HI alone,
and a total including ionized gas of $\dot{M}_{\rm acc, HVC} \ga 5 \ M_\sun {\rm \ yr^{-1}}$.
However, it is unclear if the bulk of the MS can reach the disk without being disrupted \citep{Fox14, DOnghia16}.
At present, at least a small part of it may be accreting,
as some HVCs in the Leading Arm are known to be interacting with the disk at $R \sim 17 \ {\rm kpc }$ \citep{McClure08}.
Taking the higher value of $\dot{M}_{\rm acc, HVC}$ by \cite{Lehner11},
${\cal L}_{\rm acc, HVC}$ can be up to $\sim 5$ \% of ${\cal L}_{\rm SN}$.
If the bulk of the MS can contribute to ${\cal L}_{\rm acc, HVC}$, this number may increase to $\sim 15$ \%.

The threshold mass for HVC survival in the hot halo is also uncertain.
While $M_{\rm HVC} \ga 10^{4.5} M_\sun$ is suggested from hydrodynamical simulations
for evading disruption via Kelvin-Helmholtz and Rayleigh-Taylor instabilities
\citep{Heitsch09, Kwak11, Joung12cloud},
such effects may be alleviated by inclusion of magnetic fields \citep{McCourt15},
thermal conduction \citep{Armillotta17}, or gravitational confinement by dark matter \citep{Nichols09, Galyardt16}.
If direct accretion with high $v_{\rm acc}$ is also possible for $M_{\rm HVC} \la 10^{4.5} M_\sun$,
$N_s$ can be potentially much larger than estimated by Eq. \ref{eq:N_s}.
The same is true if we account for shocks in the radiative phase at $\tau_s > 10^6 \ {\rm yr}$,
which may still accommodate some particle acceleration,
albeit up to lower energies compared to the adiabatic phase (\S \ref{sec:neutral}).

\section{Particle acceleration}
\label{sec:accel}

Shock waves are capable of accelerating charged particles to energies far exceeding their thermal values
with a power-law energy distribution,
as attested by a wide variety of observations, ranging from solar coronal mass ejections to merging clusters of galaxies.
Although its theory is far from complete, in general terms, 
DSA is expected to operate efficiently under the following conditions
(for reviews, see e.g. \citealt{Blandford87, Malkov01, Blasi13, Caprioli15}).\\
1. The relevant medium is sufficiently rarefied and magnetized so that shocks form in a collisionless manner,
mediated by collective electromagnetic interactions involving plasma instabilities rather than particle collisions (\S \ref{sec:mag}).\\
2. The magnetic field around the shock is sufficiently strong and turbulent over a range of scales
so that the injected particles can scatter back and forth across the shock front repeatedly to attain suitably high energies
within the available time and/or space constraints (\S \ref{sec:acceltime}).\\
3. An adequate fraction of particles are injected into the acceleration process,
usually requiring that either the shock has a sufficiently high Mach number,
or there is a sufficiently high density of pre-existing cosmic rays (\S \ref{sec:acceleff}).\\
4. The medium is sufficiently ionized so that collisions of ions with neutral particles do not damp such magnetic turbulence (\S \ref{sec:neutral}).

\subsection{Magnetic fields}
\label{sec:mag}

The properties of magnetic fields at the shocks of our interest are uncertain, for both RS in HVCs or FS in the outer Galactic disk.
Models of the Galactic magnetic field that are consistent with observed Faraday rotation measures
and Galactic synchrotron emission (for reviews, see \citealt{Haverkorn15, Beck16})
suggest total field strengths $B \sim 1 - 3 \ \mu$G for the disk around $R \sim 15$ kpc,
with appreciable variations depending on the exact location relative to the spiral arms
(e.g. \citealt{Jansson12, BeckM16, Planck16}).
Only 2 HVCs have published measurements of magnetic fields via Faraday rotation at levels of $B \sim 6 - 8 \ \mu$G,
subject to a number of assumptions concerning the foreground, magnetic field geometry, ionized gas distribution, etc.
\citep{McClure10, Hill13}.
These particular HVCs may be rather anomalous in having likely passed through the disk \citep{McClure10}
or possessing an unusually high metallicity \citep{Fox16}, pointing to the possibility that
more typical HVCs have weaker fields that are not readily detectable with existing instrumentation.
The general presence of metals in HVCs indicate that they are likely magnetized at some level,
although perhaps less so than in the disk due to the lack of Galactic dynamo effects.

Shock compression enhances the magnetic field components perpendicular to the shock normal by a factor $r_c$,
so if the pre-shock magnetic field with strength $B_0$ is randomly and isotropically tangled,
the post-shock field strength $B_s$ is higher by a factor $[(2r_c^2+1)/3]^{1/2}$,
whereas if it is coherent, this factor will depend on its orientation with respect to the shock.
Furthermore, magnetic fields around the shock
may possibly be amplified up to near-equipartition values by CR-induced instabilities
if the shock's Alfv\'enic Mach number ${\cal M}_A$ is sufficiently high \citep{Bell04, Caprioli15},
where ${\cal M}_A = v_s / v_A \simeq 11 \ (\mu / 0.61)^{1/2} (v_s / 300 \ {\rm km \ s^{-1}}) (B_0 / 3 {\rm \mu G})^{-1} (n_0 / 0.1 \ {\rm cm^{-3}})^{1/2}$,
and $v_A = B_0/(4 \pi \mu m_p n_0)^{1/2}$ is the Alfv\'en velocity in the pre-shock medium.
Although observationally supported for SNR shocks \citep{Uchiyama07},
the efficiency of such instabilities is unclear for the lower velocity shocks of interest here.
In our context, the strength of magnetic fields in equipartition with the CR pressure at the shock can be estimated by
$B_{s, \rm eq}^2/8\pi = (1/2) \xi_p m_p n_0 v_s^2$ so that
$B_{s, \rm eq} \simeq 14 {\rm \mu G} \ (\xi_p / 0.1)^{1/2} (n_0 / 0.1 \ {\rm cm^{-3}})^{1/2} (v_s / 300 \ {\rm km \ s^{-1}})$,
where $\xi_p$ is the fraction of kinetic energy flowing into the shock that is imparted to CR protons
(see \S \ref{sec:acceleff} for more details).
Given all these uncertainties, we take the post-shock field strength as a parameter
with a fiducial value of $B_s = 3 \ \mu {\rm G}$,
bearing in mind that either lower values or higher values up to a limit of $B_{s, \rm eq}$ may be possible.

With such values of the magnetic field around the shock,
it is straightforward to show that for protons carrying the bulk of the pre-shock kinetic energy,
the thermalization time via Coulomb collisions is much longer than the gyration timescale
over which thermalization via electromagnetic instabilities is believed to be effective \citep{Draine11}.
Condition 1 (collisionless shock) is thus satisfied.

\subsection{Acceleration time and maximum energy}
\label{sec:acceltime}

The gyroradius of a relativistic particle with charge $q$ and energy $E$ is
$r_g = E/qeB \simeq 3.6 \times 10^{-4} \ {\rm pc} \ (E/q / {\rm TeV}) (B / 3 \mu {\rm G})^{-1}$.
In terms of $v_s$ and $B_s$, the timescale for DSA of particles up to $E$ can be expressed
\begin{multline}
  \tau_{\rm DSA} \approx {10 \over 3} {\eta c E \over q e B_s v_s^2}
       \simeq 3.9 \times 10^3 \ {\rm yr} \\
       \times \eta \left({E/q \over 1 \ {\rm TeV}}\right) \left(B_s \over 3 \ \mu {\rm G}\right)^{-1} \left(v_s \over 300 \ {\rm km \ s^{-1}}\right)^{-2} ,
  \label{eq:t_accel}
\end{multline}
where $\eta \ge 1$ is a factor that depends on $r_c$, the geometry of magnetic fields on large scales,
and the amplitude and spectrum of magnetic turbulence on small scales.
For SNR shocks, observations point to $\eta \sim 1$ \citep[e.g.][]{Uchiyama07},
corresponding to Bohm-limit diffusion in fully turbulent fields.
This can be a natural consequence of the resonant streaming instability induced by the CRs \citep{Bell78},
not necessarily involving magnetic field amplification.
We assume that $\eta$ is not significantly larger than unity,
so that condition 3 (turbulent magnetic fields) is not too far from optimal for DSA.
The DSA timescale for magnetic fields in the range $B_s = 0.3 - 10 \ \mu {\rm G}$ as a function of $v_s$ is shown in Fig. \ref{fig:time}.
The maximum energy attained by protons or ions during the constant velocity phase can be evaluated by $\tau_{\rm acc} = \tau_s$,
\begin{multline}
  E_{q, \max} \approx {3 \over 10} {q e B_s v_s^2 \tau_s \over \eta c} \\
       \simeq 256 \ {\rm TeV} \ {q \over \eta} \left({B_s \over 3 \ \mu {\rm G}} {\tau_s \over 10^6 \ {\rm yr}}\right)
       \left(v_s \over 300 \ {\rm km \ s^{-1}}\right)^2 .
  \label{eq:E_qmax}
\end{multline}

In contrast, acceleration of electrons is limited by radiative losses on a timescale
\begin{multline}
  \tau_{e, {\rm rad}} = (\tau_{\rm syn}^{-1} + \tau_{\rm IC}^{-1})^{-1} = {3 m_e^2 c^3 \over 4 \sigma_T (u_B + u_{\rm ph})E_e} \\
  \simeq 5.2 \times 10^5 \ {\rm yr} \left(E_e \over {\rm TeV}\right)^{-1} \left(u_B + u_{\rm ph} \over 0.6 \ {\rm eV \ cm^{-3}}\right)^{-1} ,
  \label{eq:t_erad}
\end{multline}
where $\tau_{\rm syn}=3 m_e^2 c^3 / 4 \sigma_T u_B E_e$ and $\tau_{\rm IC}=3 m_e^2 c^3 / 4 \sigma_T u_{\rm ph} E_e$
are respectively the energy loss times of an electron with energy $E_e$
due to synchrotron and Thomson-regime inverse Compton (IC) radiation, 
$c$ is the speed of light, $m_e$ is the electron mass, $\sigma_T$ is the Thomson cross section,
$u_B = B^2 / 8\pi = 0.224 \ {\rm eV \ cm^{-3}} (B_s / 3 \ {\rm \mu G})^2$,
and $u_{\rm ph}$ is the energy density of IC seed photons \citep{Rybicki79}.
Their maximum energy is given by $\tau_{\rm acc} = \tau_{e, {\rm rad}}$ with $q=1$,
\begin{multline}
  E_{e, \max} \approx {3 m_e c v_s \over 2} \left({e B_s \over 10 \eta (u_B + u_{\rm ph})}\right)^{1/2} 
       \simeq 11.5 \ {\rm TeV} \\
       \times \eta^{-1/2} \left(B_s \over 3 \ \mu {\rm G}\right)^{1/2}  \left(u_B + u_{\rm ph} \over 0.6 \ {\rm eV \ cm^{-3}}\right)^{-1/2}
       \left(v_s \over 300 \ {\rm km \ s^{-1}}\right) .
  \label{eq:E_emax}
\end{multline}
Note that $u_{\rm ph} = u_{\rm CMB}+u_{\rm ISRF}$ comprises contributions from
the cosmic microwave background (CMB) with $u_{\rm CMB} \simeq 0.26 \ {\rm eV \ cm^{-3}}$,
and the interstellar radiation field (ISRF) with $u_{\rm ISRF}$.
The latter depends on the location in the Galaxy
and is expected to be $u_{\rm ISRF} \simeq 0.1 \ {\rm eV \ cm^{-3}}$ at $R \simeq \ 15 \ {\rm kpc}$
(\citealt{Porter08}, \S \ref{sec:dark}).
Although part of the IC losses with the ISRF may actually be in the Klein-Nishina regime (\S \ref{sec:emi_e}),
the associated modification to Eq. \ref{eq:E_emax} should be minor,
due to the persistence of Thomson-regime IC losses with the CMB as well as synchrotron losses.

\begin{figure}[!htb]
\epsscale{1.2}
\plotone{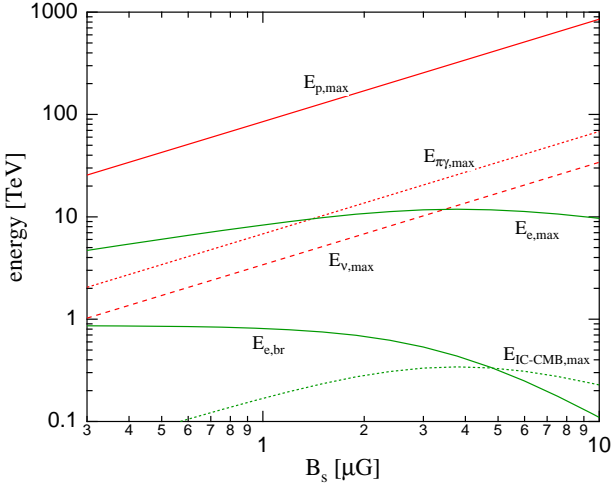}
\caption{
Characteric particle and photon energies versus post-shock magnetic field $B_s$.
Compared are maximum energies of protons $E_{p, \max}$ (Eq. \ref{eq:E_qmax} with $q=1$; red solid line)
and electrons $E_{e, \max}$ (Eq. \ref{eq:E_emax}; upper green solid curve),
assuming an interstellar radiation field with $u_{\rm ISRF} \simeq 0.1 \ {\rm eV \ cm^{-3}}$ as appropriate for $R \simeq \ 15 \ {\rm kpc}$.
Also shown are the maximum energies of $pp$ collision-induced
gamma-rays $E_{\pi \gamma, \max} = 0.08 \ E_{p, \max}$ (red dashed line)
and neutrinos $E_{\nu, \max} = 0.04 \ E_{p, \max}$ (red long-dashed line),
the electron cooling break energy $E_{e, {\rm br}}$ (Eq. \ref{eq:E_ebr}; lower green solid curve),
and the maximum energy of inverse Compton photons via upscattering
of the cosmic microwave background $E_{\rm IC-CMB,\max}$ (Eq: \ref{eq:E_ICmax}; green dashed curve).
}
\label{fig:energy}
\end{figure}

Fig. \ref{fig:energy} compares $E_{p, \max}$ (Eq. \ref{eq:E_qmax} with $q=1$) and $E_{e, \max}$
as functions of $B_s$ in the range $0.3 - 10 {\rm \mu G}$.
Note that in constrast to $E_{p, \max}$ that increases monotonically in proportion to $B_s$,
the dependence of $E_{e,\max}$ on $B_s$ is much weaker.
Moreover, $E_{e,\max}$ increases with $B_s$ up to $E_{e,\max} \simeq 12 \ {\rm TeV}$ at $B_s \sim 3.8 {\rm \mu G}$,
but decreases beyond due to the onset of synchrotron-dominant losses.

\subsection{Acceleration efficiency and spectrum}
\label{sec:acceleff}

The acceleration efficiency, i.e. the fraction of energy that is channeled into accelerated particles by the DSA mechanism,
is determined by physical processes that are not yet fully understood.
For accelerated protons, various observations and some numerical simulations indicate that
their total energy can be up to $\sim$ 10 - 20 \% of the pre-shock kinetic energy,
as long as ${\cal M}_s \ga 10$
 (\citealt{Ackermann13, Kang13, Caprioli15} and references therein).
As discussed in \S \ref{sec:shockvt}, the latter should be valid in most cases for the shocks considered here,
justifying condition 2 (adequate shock Mach number).
Also governed by ${\cal M}_s$ is the spectrum of accelerated particles,
expected to be a simple power law in particle momentum $p$ to first order of approximation,
$dN/dp \propto p^{-\alpha}$, with index $\alpha=(r_c+2)/(r_c-1)=(3\gamma-1+4{\cal M}_s^{-2})/(2-2{\cal M}_s^{-2})$
that approaches $\alpha = 2$ for ${\cal M}_s \ga 10$.
We assume that during the pre-deceleration, constant velocity phase of the shock with duration $\tau_s$,
non-thermal protons with momenta $p \ge p_{p,\min}$
are injected into the emission region at a constant rate
\begin{equation}
  Q_p(p) \equiv {dN_p \over dp dt} = Q_{p0} \left(p \over p_0\right)^{-\alpha} \exp \left(- {p \over p_{p,\max}}\right)
   \label{eq:Q_p}
\end{equation}
where $Q_p(p)dp$ denotes the number of protons with momenta in the interval $p \sim p+dp$ injected per unit time,
$p_0 = 1 \ {\rm GeV/c}$, and $\alpha=2$ is fiducially considered.
The minimum momentum of protons is expected to be not far above their thermal values,
for which we choose $p_{p,\min} = 0.01 \ {\rm GeV/c}$,
while $p_{p,\max} = E_{p,\max}/c$ is evaluated from Eq. \ref{eq:E_qmax}.
The normalization parameter $Q_{p0}$ is treated differently depending on the objective, 
in relation to the total power of injected protons $L_p = \int_{p_{p,\min}}^{p_{p,\max}} E_k Q_p(p) dp$
where $E_k = m_p c^2 (\sqrt{((p/m_p c)^2+1)}-1)$ is the proton kinetic energy,
or the time-integrated total proton energy $W_p \approx L_p \tau_s$.
For giving predictions, $Q_{p0}$ can be set so that $L_p = \xi_p L_{k, {\rm HVC}}$,
proportional to the rate of HVC kinetic energy passing through the shock (Eq. \ref{eq:L_kin}, \S \ref{sec:enenum}),
with fiducial value $\xi_p = 0.1$.
On the other hand, for providing model fits to observations, $Q_{p0}$ can be adjusted to give the best description of the data,
and the corresponding $L_p$ or $W_p$ is evaluated a posteriori for a plausibility check.
We do not consider effects that can induce deviations from a simple power-law for the accelerated particle spectrum
such as non-linear feedback from CRs onto the shock structure \citep{Malkov01},
nor the effects of pre-existing CRs for DSA injection, which is likely subdominant for the conditions of our interest.

Primary electrons, i.e. those directly accelerated out of the thermal plasma via DSA,
are treated in a way similar to protons (Eq. \ref{eq:Q_p}) with the same value of $\alpha$ so that their injection rate
with momenta  $p_e \ge p_{e,\min}$ is
 \begin{equation}
  Q_e(p_e) \equiv {dN_e \over dp_e dt} = Q_{e0} \left(p_e \over p_0\right)^{-\alpha} \exp \left(- {p \over p_{e,\max}}\right)
  \label{eq:Q_e}
\end{equation}
where $p_{e,\max} = E_{e,\max}/c$ is evaluated from Eq. \ref{eq:E_emax}.
The minimum momentum of electrons is uncertain but its exact value is not crucial for our purposes
as long as $p_{e,\min} \ll 1 {\rm GeV/c}$, where the corresponding synchrotron and IC emission
is observationally irrelevant (\S \ref{sec:emi_e}); here we take $p_{e,\min} = 0.01 \ {\rm GeV/c}$.

The electron acceleration efficiency $\xi_e$, defined so that the total injected electron power $L_e = \xi_e L_{k, {\rm HVC}}$,
is more uncertain compared to protons, either observationally or theoretically.
Often invoked for the ratio of accelerated electrons to protons 
is $K_{ep} \simeq 0.01$ in terms of their number at momentum $\sim 1 {\rm GeV/c}$,
the value observed in Galactic CRs.
However, it is unclear how much this reflects the ratio at the acceleration site.
Observations and related simulations for SNRs indicate that it can be much less \citep{Ackermann13, Caprioli15},
while those for cluster merger shocks suggest that it may be appreciably higher \citep{Guo14}.
We fiducially take $K_{ep} = Q_{e0} / Q_{p0} = 0.01$,
but remain open to significantly different values.

\subsection{Neutral particles and radiative regime}
\label{sec:neutral}

If the medium around the shock contains a sufficiently large fraction of neutral particles,
their collisions with ions can damp the magnetic turbulence that is essential for the DSA process and curtail it \citep{Bell78, Drury96}.
For both RS in HVCs and FS in the disk, a major fraction of the pre-shock gas should be HI.
On the other hand, irrespective of the pre-shock ionization state,
shocks in the radiative regime can fully ionize the upstream gas
via UV photons from the downstream gas, as long as $v_s \ga 120 {\rm \ km \ s^{-1}}$ \citep{Shull79, Hollenbach89}.
While our main concern is the adiabatic phase of the shocks,
$\tau_{g, {\rm rad}}$ was seen to be comparable to $\tau_{\rm HVC}$ or $\tau_{\rm disk}$ with our fiducial parameters (\S \ref{sec:shockvt}).
Thus, unless we consider the earliest phases of their evolution, our shocks are expected to be at least moderately radiative,
likely substantially mitigating the effects of neutral particles.
Here we assume the validity of condition 4 (negligible damping of magnetic turbulence by neutral-ion collisions),
and defer a more detailed discussion to the future.
For the same reason, we do not consider in this work the effects of charge exchange reactions that may lead to non-trivial consequences
\citep[e.g.][]{Ohira12, Morlino13}.

We note that particle acceleration may also continue in the radiative phase,
possibly with a spectral break above a few GeV when $v_s \la 120 {\rm \ km \ s^{-1}}$,
as proposed in order to explain some observations of old SNRs and other objects
\citep[e.g.][]{Bykov00, Yamazaki06, Malkov11, Lee15, InoueY17}.
However, such effects are not yet understood in detail and will not be discussed here.

\section{Non-thermal emission}
\label{sec:emission}

For detailed calculations of non-thermal emission and application to unidentified Galactic GeV-TeV sources in \S \ref{sec:HVCgam},
we employ a numerical code used in \citet{Uchiyama10},
based on a time-dependent kinetic description of the non-thermal proton and electron populations
within a suitable emission region.
For protons, the code accounts for their inelastic collisions with ambient matter and consequent pion production,
emission due to decay of neutral pions ($\pi^0$),
and injection of secondary electrons and positrons ($e^\pm$; hereafter simply ``secondary electrons'' unless otherwise noted)
due to decay of charged pions ($\pi^\pm$).
The contribution of helium and other heavy nuclei is accounted for by the nuclear enhancement factor
\citep{Mori09, Kachelriess14},
for which we adopt $\epsilon_M = 2.0$ (see \S \ref{sec:emi_p} for more details).
For both primary and secondary electrons, the code includes bremsstrahlung with ambient matter,
synchrotron emission in magnetic fields, and inverse Compton emission by upscattering the CMB and ISRF
with proper account of the Klein-Nishina regime.
For simplicity, the ISRF spectrum is described by two diluted black-body components
with temperature and normalization for each
chosen so as to approximate the detailed calculations by \cite{Porter08}
of the optical starlight and far-infrared dust emission that depend on $R$.
The proton and electron distributions are calculated self-consistently in a time-dependent way
including the effect of energy losses due to all of the above processes.

Below we provide a simplified discussion of the key emission processes,
for the sake of estimates and understanding of the numerical results.
Readers familiar with the basics of non-thermal emission may skip the rest of this section
and move to \S \ref{sec:HVCgam}.

\subsection{Emission induced by protons}
\label{sec:emi_p}

Inelastic collisions of relativistic protons with stationary protons and nuclei in ambient matter
lead to production of mainly pions, partitioned roughly equally among $\pi^0$, $\pi^+$ and $\pi^-$,
of which the $\pi^0$ decay to gamma-rays, and the $\pi^\pm$ decay to neutrinos and $e^\pm$.
For a proton with sufficiently high energy $E_p$, the mean energy of the produced gamma rays and neutrinos
are $E_\gamma \approx 0.08 \ E_p$ and $E_\nu \approx 0.04 \ E_p$, respectively.
The total inelastic $pp$ cross section at $E_p > 10$ GeV can be approximated by
$\sigma_{pp} \simeq 3 \times 10^{-26} \ {\rm cm^2} \ [0.95 + 0.06 \ln(E_p/{\rm GeV})]$,
which increases logarithmically with $E_p$ \citep{Aharonian04, Dermer09}.
The relevant energy loss timescale for a proton with $E_p \sim 100 \ {\rm TeV}$ is
\begin{multline}
  \tau_{pp} \approx (n \kappa_{pp} \sigma_{pp} c)^{-1} = 1.1 \times 10^8 \ {\rm yr} \left({r_c n_0 \over 0.4 \ {\rm cm^{-3}}} \right)^{-1} ,
  \label{eq:t_pp}
\end{multline}
where $\kappa_{pp} \simeq 0.5$
is the average inelasticity of the $pp$ collision,
and the main target matter is considered to be the post-shock gas with density $n = r_c n_0$.
The resulting gamma-ray spectrum rises with $E_\gamma$ abruptly
at $E_\gamma \la E_\pi = m_\pi c^2 / 2 \simeq 67.5 \ {\rm MeV}$ where $m_\pi$ is the $\pi^0$ mass,
mirrors that of the parent protons with similar spectral index,
$d{\cal N}_{\gamma,pp}/dE_\gamma \propto E_\gamma^{-\alpha}$,
at $E_\pi \la E_\gamma \la E_{\pi \gamma,\max} \approx 0.08 \ E_{p,\max}$, and cuts off at $E_\gamma \ga E_{\pi \gamma,\max}$.

At relativistic energies, the proton injection rate is well described by $dN_p/dE_p dt \propto E_p^{-\alpha}$.
When $\alpha \simeq 2$, the total proton power can be approximated by
$L_p = \int_{E_{p,\min}}^{E_{p,\max}} E_p (dN_p/dE_p dt) dE_p$,
with $E_{p,\min} \simeq 1 \ {\rm GeV}$ and $E_{p,\max}$ given by Eq. \ref{eq:E_qmax}. 
Assuming no escape of protons from the emission region,
an estimate for the gamma-ray energy flux around $E_\gamma \sim 8 \ {\rm TeV}$
from a source at distance $D$ is \citep[e.g.][]{Drury94, Murase08}
\begin{multline}
  E_\gamma^2 {d{\cal N}_{\gamma,pp} \over dE_\gamma} \approx {1 \over 3} E_p^2 {dN_p \over dE_p dt} {\epsilon_M \min[1,f_{pp}] \over 4 \pi D^2} \\
  \simeq 3.2 \times 10^{-12} \ {\rm erg \ cm^{-2} s^{-1}}
  \left({C_p \over 0.08} {\xi_p L_{k, {\rm HVC}} \over 3.1 \times 10^{38} \ {\rm erg \ s^{-1}}} {\tau_s \over 10^6 \ {\rm yr}}\right) \\
  \times \left({\epsilon_M \over 2.0} {r_c n_0 \over 0.4 \ {\rm cm^{-3}}} \right)
  \left(D \over 20 \ {\rm kpc}\right)^{-2} ,
  \label{eq:flux_pp}
\end{multline}
where $f_{pp} = \tau_s / \tau_{pp} \simeq 0.0093 (r_c n_0/ 0.4 \ {\rm cm^{-3}})(\tau_s/ 10^6 \ {\rm yr})$ at $E_p \simeq 100 \ {\rm TeV}$,
$C_p = \ln (E_{p,\max}/E_{p,\min})^{-1}$,
and the numerical expression is for our fiducial parameters with $\alpha=2$
and $E_{p,\max} \simeq 256 \ {\rm TeV}$ (Eq. \ref{eq:E_qmax}).
The flux in the range $E_\pi \la E_\gamma \la E_{\gamma,\max}$ should be similar for $\alpha=2$.
Such values are typical for the numerous TeV sources found in the H.E.S.S. Galactic Plane Survey (\S \ref{sec:unID}).

The nuclear enhancement factor $\epsilon_M$ takes into account
the additional contribution to pion production by He and heavier nuclei, both in the CRs and in the target matter.
For Galactic CRs impinging on gas with solar composition, \cite{Mori09} gives
$\epsilon_M = 1.84$ for CR kinetic energy of 10 GeV/nucleon, slowly increasing to $\epsilon_M = 2.00$ for 1 TeV/nucleon.
In our case, the main energy of interest is still higher,
while the abundance of nuclei heavier than C are likely much lower in the CRs and target gas.
Further uncertainties concern details of hadronic interactions \citep{Kachelriess14}
and relative spectral variance among CR species near the source that may differ from Galactic CRs.
In view of the ambiguities, we choose $\epsilon_M = 2.0$ for simplicity,
noting that a precise value is not critical for our aims.

The flux of co-produced neutrinos per flavor at $E_\nu \approx 0.04 \ E_p$ after neutrino oscillations
is approximately 1/2 of Eq. \ref{eq:flux_pp}.
Considering $E_p \sim 250 \ {\rm TeV}$, the flux at $E_\nu \simeq 10 \ {\rm TeV}$ is
\begin{multline}
  E_\nu^2 {d{\cal N}_{\nu,pp} \over dE_\nu} \approx {1 \over 6} E_p^2 {dN_p \over dE_p dt} {\epsilon_M \min[1,f_{pp}] \over 4 \pi D^2} \\
  \simeq 1.8 \times 10^{-12} \ {\rm erg \ cm^{-2} s^{-1}}
  \left({C_p \over 0.08} {\xi_p L_{k, {\rm HVC}} \over 3.1 \times 10^{38} \ {\rm erg \ s^{-1}}} {\tau_s \over 10^6 \ {\rm yr}}\right) \\
  \times \left({\epsilon_M \over 2.0} {r_c n_0 \over 0.4 \ {\rm cm^{-3}}} \right)
  \left(D \over 20 \ {\rm kpc}\right)^{-2} ,
  \label{eq:flux_nu}
\end{multline}
where the numerical expression is for our fiducial parameters with $\alpha=2$.
This may be beyond the capability of current neutrino observatories such as IceCube,
but within reach of future facilities (\S \ref{sec:neuobs}).

\subsection{Emission induced by electrons}
\label{sec:emi_e}

Primary electrons of sufficiently high energy can cool efficiently within the available time due to synchrotron and IC losses,
leading to a break in their energy distribution relative to $dN_{e}/dE_e \propto E_e^{-\alpha}$
above the energy where $\tau_{e, {\rm rad}} = \tau_s$,
\begin{multline}
  E_{e,{\rm br}} \approx {3 m_e^2 c^3 \over 4 \sigma (u_B + u_{\rm ph}) \tau_s}
       \simeq 0.52 \ {\rm TeV} \left({u_B + u_{\rm ph} \over 0.6 \ {\rm eV \ cm^{-3}}} {\tau_s \over 10^6 \ {\rm yr}}\right)^{-1} ,
  \label{eq:E_ebr}
\end{multline}
where the expression for IC losses in the Thomson regime has been used (\S \ref{sec:acceltime}).
If $E_{e,{\rm br}} > E_{e,\max}$, $dN_{e}/dE_e \propto E_e^{-\alpha}$ is maintained up to $E_e \sim E_{e,\max}$.
Fig. \ref{fig:energy} shows $E_{e,{\rm br}}$ as a function of $B_s$ when $u_{\rm ISRF} \simeq 0.1 \ {\rm eV \ cm^{-3}}$.

An electron with given $E_e$ radiates synchrotron photons with characteristic energy
$E_{\rm sy} = \epsilon_B (E_e/m_e c^2)^2$, where
$\epsilon_B = 3 h e B/4 \pi m_e c^2$ and $h$ is the Planck constant \citep{Rybicki79}.
When $E_{e,{\rm br}} < E_{e,\max}$,
the synchrotron spectrum is characterized by
$d{\cal N}_{\gamma, {\rm sy}}/dE_\gamma \propto E_\gamma^{-(\alpha+1)/2}$,
becoming steeper at $E_{\rm sy,br} \la E_\gamma \la E_{\rm sy,\max}$
and cutting off at $E_\gamma \ga E_{\rm sy,\max}$,
where
$E_{\rm sy,br} = \epsilon_B (E_{e,{\rm br}}/m_e c^2)^2$
and
\begin{multline}
  E_{\rm sy,\max} = \epsilon_B (E_{e,\max}/m_e c^2)^2 \\
  \simeq 26 \ {\rm eV} \ \eta^{-1} \left({B_s \over 3 \ \mu {\rm G}} {v_s \over 300 \ {\rm km \ s^{-1}}}\right)^2 \left(u_B + u_{\rm ph} \over 0.6 \ {\rm eV \ cm^{-3}}\right)^{-1} .
  \label{eq:E_symax}
\end{multline}
As $B_s < B_{s, \rm eq} \simeq 14 {\rm \mu G}$ (\S \ref{sec:mag}), the primary synchrotron spectrum cannot extend beyond the far UV band,
implying that HVC accretion events may not be readily detectable in X-rays (\S \ref{sec:results}, \S \ref{sec:Xradobs}).
This is in stark contrast to SNRs or PWNe for which $E_{\rm sy,\max}$
is generally at X-ray energies or higher, due to their much higher shock velocities and/or magnetic fields.

Upscattering of seed photons with energy $\epsilon_0$ by an electron with given $E_e$
results in IC photons with characteristic energy $E_{\rm IC} = \epsilon_0 (E_e/m_e c^2)^2$ 
when it occurs in the Thomson regime with $E_e \ll (m_e c^2)^2/ \epsilon_0$. 
This corresponds to $E_e \ll 410 \ {\rm TeV}$ if the seeds are the CMB
with typical photon energy $\epsilon_0 = \epsilon_{\rm CMB} \simeq 6.4 \times 10^{-4} \ {\rm eV}$,
or $E_e \ll 0.26 \ {\rm TeV}$ if they are the starlight component of the ISRF
with $\epsilon_0 = \epsilon_{\rm star} \simeq 1 \ {\rm eV}$.
When $E_e \ga (m_e c^2)^2/ \epsilon_0$,
scattering proceeds in the Klein-Nishina regime where the relation between $E_{\rm IC}$ and $E_e$ is not as simple,
other than the limit $E_{\rm IC} < E_e$.
With our fiducial parameters, electrons with $E_e = E_{e,{\rm br}}$ are in the Thomson regime for either CMB or ISRF
so that the associated IC photon energy is $E_{\rm IC,br} = \epsilon_0 (E_{e,{\rm br}}/m_e c^2)^2$.
On the other hand, electrons with $E_e = E_{e,\max}$ are in the Thomson regime only for the CMB,
for which the IC photon energy
\begin{multline}
  E_{\rm IC-CMB,\max} = \epsilon_{\rm CMB} (E_{e,\max}/m_e c^2)^2 \\
   \simeq 0.32 \ {\rm TeV} \eta^{-1}
   \left({B_s \over 3 \ \mu {\rm G}}\right) \left({v_s \over 300 \ {\rm km \ s^{-1}}}\right)^2 \left(u_B + u_{\rm ph} \over 0.6 \ {\rm eV \ cm^{-3}}\right)^{-1} ,
  \label{eq:E_ICmax}
\end{multline}
which is plotted in Fig. \ref{fig:energy} as a function of $B_s$ when $u_{\rm ISRF} \simeq 0.1 \ {\rm eV \ cm^{-3}}$.
For the ISRF starlight, such electrons are in the Klein-Nishina regime instead,
so that $E_{\rm IC-ISRF,\max}$ is limited by $E_{e,\max} \la 12 \ {\rm TeV}$ (\S \ref{sec:acceltime}, Fig. \ref{fig:energy}).
When $E_{e,{\rm br}} < E_{e,\max}$,
the contribution to the IC spectrum from seed photons with given $\epsilon_0$
is analogous to that for synchrotron,
with $d{\cal N}_{\gamma,{\rm IC}}/dE_\gamma \propto E_\gamma^{-(\alpha+1)/2}$ at $E_\gamma \la E_{\rm IC,br}$,
becoming steeper at $E_{\rm IC,br} \la E_\gamma \la E_{\rm IC,\max}$ and cutting off at $E_\gamma \ga E_{\rm IC,\max}$.
The total IC spectrum is represented by the convolution over the CMB and ISRF seed spectra,
resulting in broader break and cutoff features as well as a steeper cutoff due to Klein-Nishina effects.

At relativistic energies, the injection rate for electrons is $K_{ep}$ times that for protons evaluated at the same energy,
$(dN_e/dE_e dt) |_{E_e = E} = K_{ep} (dN_p/dE_p dt) |_{E_p = E}$.
The steady-state distribution of electrons at $E_e \la E_{e,{\rm br}}$ should be roughly unchanged from the injected distribution,
$\tau_s (dN_e/dE_e dt)$.
Considering only the CMB as seed photons, the IC energy flux at $E_\gamma < E_{\rm IC,br}$
can be estimated by substituting $E_e=(E_\gamma/\epsilon_{\rm CMB})^{1/2} m_e c^2$ in
\begin{multline}
  E_\gamma^2 {d{\cal N}_{\gamma,{\rm IC}} \over dE_\gamma} \approx {1 \over 2} E_e^2 {dN_e \over dE_e dt}
  {\tau_s \over \tau_{\rm IC}(E_e)} {1 \over 4 \pi D^2} \\
  \simeq 1.4 \times 10^{-12} \ {\rm erg \ cm^{-2} s^{-1}} \left(E_\gamma \over {\rm GeV}\right)^{1/2} \qquad \qquad \qquad \\
  \qquad \times \left({K_{ep} \over 0.01} {C_p \over 0.08} {\xi_p L_{k, {\rm HVC}} \over 3.1 \times 10^{38} \ {\rm erg \ s^{-1}}} {\tau_s \over 10^6 \ {\rm yr}}\right)
  \left(D \over 20 \ {\rm kpc}\right)^{-2}
  \label{eq:flux_IC}
\end{multline}
\citep[e.g.][]{Dermer09},
where the numerical expression is for our fiducial parameters with $\alpha=2$,
somewhat less than the estimated $\pi^0$ gamma-ray flux (Eq. \ref{eq:flux_pp}).

Although the above estimate is for primary electrons only,
depending on the value of $K_{ep}$,
the synchrotron and IC emission from secondary $e^{\pm}$ induced by $pp$ collisions
can be non-negligible, and in some cases even dominant over primary electrons
for certain energy bands (\S \ref{sec:results}).

\section{HVC accretion origin of high-energy gamma-ray sources}
\label{sec:HVCgam}

\subsection{Unidentified Galactic GeV-TeV Sources and HESS J1503-582}
\label{sec:unID}

Observations of the Galactic Plane in TeV gamma rays over the past decades
have revealed numerous sources
that are spatially extended on scales $\sim 0.03 - 0.3 ^\circ$
and lack obvious counterparts at other wavelengths.
After the first object found by HEGRA \citep{Aharonian02},
dozens of such sources were discovered in the inner Galactic Plane Survey (GPS) conducted by H.E.S.S.
at Galactic latitudes $|b| < 5^ \circ$ with angular resolution $\sim 0.07 ^\circ$
\citep{Aharonian05, Aharonian06, Aharonian08}.
New sources in this class are continuing to be found
in deeper and more extended surveys by H.E.S.S. \citep{Deil15, Donath17},
and in ongoing surveys by the recently completed HAWC array \citep{Abeysekara17}.
Similar sources have also been reported by
MILAGRO \citep{Abdo07}, ARGO \citep{Bartoli13},
VERITAS \citep{Weinstein09} and MAGIC \citep{Aleksic14}.
As of August 2017, TeVCat \citep{Wakely08}
lists $\sim$55 sources located within $\sim 10 ^\circ$ of the Galactic Plane with type ``unidentified''.

A fair number of unidentified sources along the Galactic Plane
have also been uncovered in all-sky surveys at GeV energies,
most recently by {\it Fermi}-LAT.
In the 2FHL catalog based on data at 50 GeV - 2 TeV with angular resolution $\sim 0.1 ^\circ$,
22 sources at $|b|< 10 ^\circ$ are listed without obvious identification \citep{Ackermann16}.
The {\it Fermi}-LAT Galactic Extended Source Catalog reveals 8 sources newly detected above 10 GeV
with extension $\ga 0.3 ^\circ$
that are not clearly associated with known objects \citep{Ackermann17_ES}.

The origin of such unidentified Galactic GeV-TeV sources has been debated,
with various proposed explanations
including old SNRs \citep{Yamazaki06}, middle-aged PWNe \citep{deJager09},
gamma-ray burst remnants \citep{Atoyan06, Ioka10}, etc.
After dedicated follow-up studies across the electromagnetic spectrum,
a major fraction of TeV sources that initially lacked identification
have later been recognized as known types of objects,
especially PWNe \citep{Abdalla17} and SNRs \citep{Gottschall16}.

A unique source that has defied clarification and remains mysterious is HESS J1503-582 \citep{Renaud08}.
Located at Galactic coordinates $(l, b) \sim (319.7 ^\circ, 0.3 ^\circ)$
and spatially extended with root mean square size $\sim 0.26 ^\circ$,
its spectrum at 1.3 - 22 TeV
can be fit by a power-law with photon index $\Gamma_{\rm HESS} = 2.4 \pm 0.6$
and flux normalization at 1 TeV of $(1.6 \pm 0.6) \times 10^{-12} \ {\rm cm^{-2} \ s^{-1} \ TeV^{-1}}$.
A counterpart at 50 GeV - 2 TeV is identified in the 2FHL catalog,
with a spectrum connecting smoothly to that measured by H.E.S.S. \citep{Ackermann16}.
On the other hand, no obvious counterparts have been found at X-ray, infrared or radio wavelengths,
nor any correlations with known classes of objects such as SNRs, HII regions, star forming regions, etc.

Intriguingly, \cite{Renaud08} discuss a potential association with the ``forbidden velocity wing'' FVW 319.8+0.3.
As defined by \cite{Kang07},
FVWs are structures observed in HI emission near the Galactic Plane at $|b| < 1.5 ^\circ$
with spatial extension $\la 2 ^\circ$ and velocity deviating from Galactic rotation by more than $\sim 20 {\rm \ km \ s^{-1}}$.
Of the 87 FVWs identified by \cite{Kang07} in large-scale HI maps
with spatial and spectral resolution of $0.5 ^\circ$ and $\sim 1 {\rm \ km \ s^{-1}}$, respectively,
a few are spatially coincident with known SNRs, HVCs, or nearby galaxies.
However, no such associations were found for 85 \% of their sample,
and the distance and nature of most FVWs are unknown.
TeVCat lists HESS J1503-582 as a ``dark'' source,
currently the only object with such a designation.

\subsection{HVC accretion origin of the dark source HESS J1503-582}
\label{sec:dark}

As introduced in \S \ref{sec:HVCacc},
one object in the sample of \cite{Kang07}, FVW 40.0+0.5, 
was recently revealed through HI observations with higher spatial and spectral resolution ($0.066 ^\circ$ and $0.184 {\rm \ km \ s^{-1}}$)
to consist of an expanding kpc-scale supershell
with a CHVC
at its geometric center \citep{Park16}.
The most natural interpretation is a HVC accretion event,
that is, the impact of the CHVC at high velocity with the Galactic disk giving rise to the supershell.
 \cite{Park16} favor a location in the outer Galaxy at distance $D \sim 20 {\rm \ kpc}$,
which corresponds to Galactocentric radius $R \sim 15 {\rm \ kpc}$ at its sky position in the first Galactic quadrant.
From the observed properties of the supershell and estimates of the required total energy,
they infer that the CHVC began its impact $\sim 5 \times 10^6$ yr ago
with initial kinetic energy $E_{\rm HVC} \ga 7 \times 10^{52} {\rm \ erg}$.
Assuming a relative collision velocity $v \sim 240 {\rm \ km \ s^{-1}}$,
this implies an initial mass $M_{\rm HVC} \ga 6 \times 10^4 \ {\rm M_\sun}$,
much larger than the currently observed HI mass $M_{\rm HI} \simeq 5800 \ {\rm M_\sun}$,
which is possibly a consequence of ram pressure stripping during the accretion process.

Bolstered by this finding, we propose that HESS J1503-582, spatially coincident with FVW 319.8+0.3,
also originated from a direct HVC accretion event in the outer Galaxy,
and apply our model as formulated in \S \ref{sec:shock} - \S \ref{sec:emission}.
For concreteness, we focus on an interpretation based on the RS within the HVC,
although a connection with the FS in the Galactic disk is not excluded.
Our most uncertain parameter is the magnetic field around the shock, for which we consider a range of possible values,
$B_s = 0.3 - 10 \ {\rm \mu G}$.
Otherwise, the fiducial parameter values proposed in \S \ref{sec:shock} and \S \ref{sec:accel} are chosen,
except for some deviations in $n_{\rm HVC}$, $\xi_p$ and $K_{ep}$
in order to provide the most consistent description of the existing observations.

For the HVC, 
we choose parameters similar to those inferred by \cite{Park16} for their pre-collision CHVC:
$v_s = 300 {\rm \ km \ s^{-1}}$,
$M_{\rm HVC} = 10^5 \ {\rm M_\sun}$,
and $n_{\rm HVC} = 0.15 \ {\rm cm^{-3}}$,
the latter implying $r_c n_{\rm HVC} = 0.6 \ {\rm cm^{-3}}$ for the post-shock gas.
The distance is taken to be $D = 20 {\rm \ kpc}$,
corresponding to an impact location in the outer Galaxy at $R = 15 {\rm \ kpc}$
for the sky position of HESS J1503-582 in the fourth Galactic quadrant,
quite analogous to the system of \cite{Park16}.
These parameters give
$\tau_{\rm HVC} \simeq 1.2 \times 10^6 \ {\rm yr}$,
$\tau_{\rm disk} \simeq 2.0 \times 10^6 \ {\rm yr}$ and
$\tau_{g, {\rm rad}}  \simeq 1.2 \times 10^6 \ {\rm yr}$,
so that the duration of the constant velocity phase is determined by $\tau_{\rm HVC}$ and/or $\tau_{g, {\rm rad}}$.
For simplicity, we assume that the event is being observed at age $t = 10^6 \ {\rm yr}$ after the beginning
of shock formation and particle acceleration, so that quantities discussed in \S \ref{sec:accel} and \S \ref{sec:emission}
can be evaluated with $\tau_s = t = 10^6 \ {\rm yr}$.
At $R = 15 \ {\rm kpc}$,
the expected ISRF \citep{Porter08}
is approximated by two diluted black-body components with temperatures and energy densities representing
the far infrared dust emission ($k_B T_{\rm dust} = 3.0 \times 10^{-3} \ {\rm eV}$, $u_{\rm dust} = 0.05 \ {\rm eV \ cm^{-3}}$)
and the optical starlight ($k_B T_{\rm star} = 0.25 \ {\rm eV}$, $u_{\rm star} = 0.05 \ {\rm eV \ cm^{-3}}$),
which total $u_{\rm ISRF} = 0.1 \ {\rm eV \ cm^{-3}}$
and is subdominant compared to the CMB
($k_B T_{\rm CMB} = 2.3 \times 10^{-4} \ {\rm eV}$, $u_{\rm CMB} = 0.26 \ {\rm eV \ cm^{-3}}$).
Concerning particle acceleration, we take $\eta=1$ and $\alpha=2$,
while tolerating large departures from $K_{ep}=0.01$ if required by the observations (\S \ref{sec:acceleff}).
Finally, the normalization of the proton distribution is adjusted to provide a viable fit to the gamma-ray data,
and the corresponding value of $\xi_p$ is checked a posteriori for plausibility.

Note that for a uniform spherical HVC, the above parameters give $r_{\rm HVC} \simeq 186 \ {\rm pc}$
and an angular diameter $\theta \approx 2 r_{\rm HVC}/D \simeq 1.1 ^\circ$,
somewhat larger than the observed rms angular diameter $0.52 ^\circ$ for HESS J1503-582.
However, this is not deemed to be an issue,
as the gas density profile of a real CHVC is likely more centrally concentrated than a uniform distribution,
not to mention the possibility of non-trivial collision geometry and projection effects, etc.

The currently available broadband data for HESS J1503-582 are shown in Figs. \ref{fig:B3muG} - \ref{fig:B03muG}.
In addition to the gamma-ray data from H.E.S.S. \citep{Renaud08} and {\it Fermi}-LAT \citep{Ackermann16},
also plotted are radio and X-ray upper limits, estimated with the methods described in \cite{Abramowski11}
and originally applied to HESS J1356-645, a source with spatial extension similar to HESS J1503-582.
Conservative upper limits from the Molonglo Galactic Plane Survey (MGPS) at 843 MHz
and the Parkes-MIT-NRAO (PMN) survey at 4.85 GHz are 0.61 Jy and 0.62 Jy, respectively.
To derive X-ray upper limits, we assume spectral index $\Gamma_X = 2$
and intervening hydrogen column density $N_{\rm H} \simeq 1.5 \times 10^{22} \ {\rm cm^{-2}}$,
which is the total Galactic value toward the direction of HESS J1503-582 \footnote{\url{https://heasarc.nasa.gov/cgi-bin/Tools/w3nh/w3nh.pl}}
and likely comparable to the actual value at our assumed distance $D=20$ kpc.
This gives a 3$\sigma$ limit from the {\it ROSAT} All-Sky Survey (RASS) in the 1-2.4 keV band
of $9 \times 10^{-12} \ {\rm erg \ cm^{-2} s^{-1}}$.

\subsection{Model results for HESS J1503-582}
\label{sec:results}

\begin{figure}[!htb]
\epsscale{1.3}
\plotone{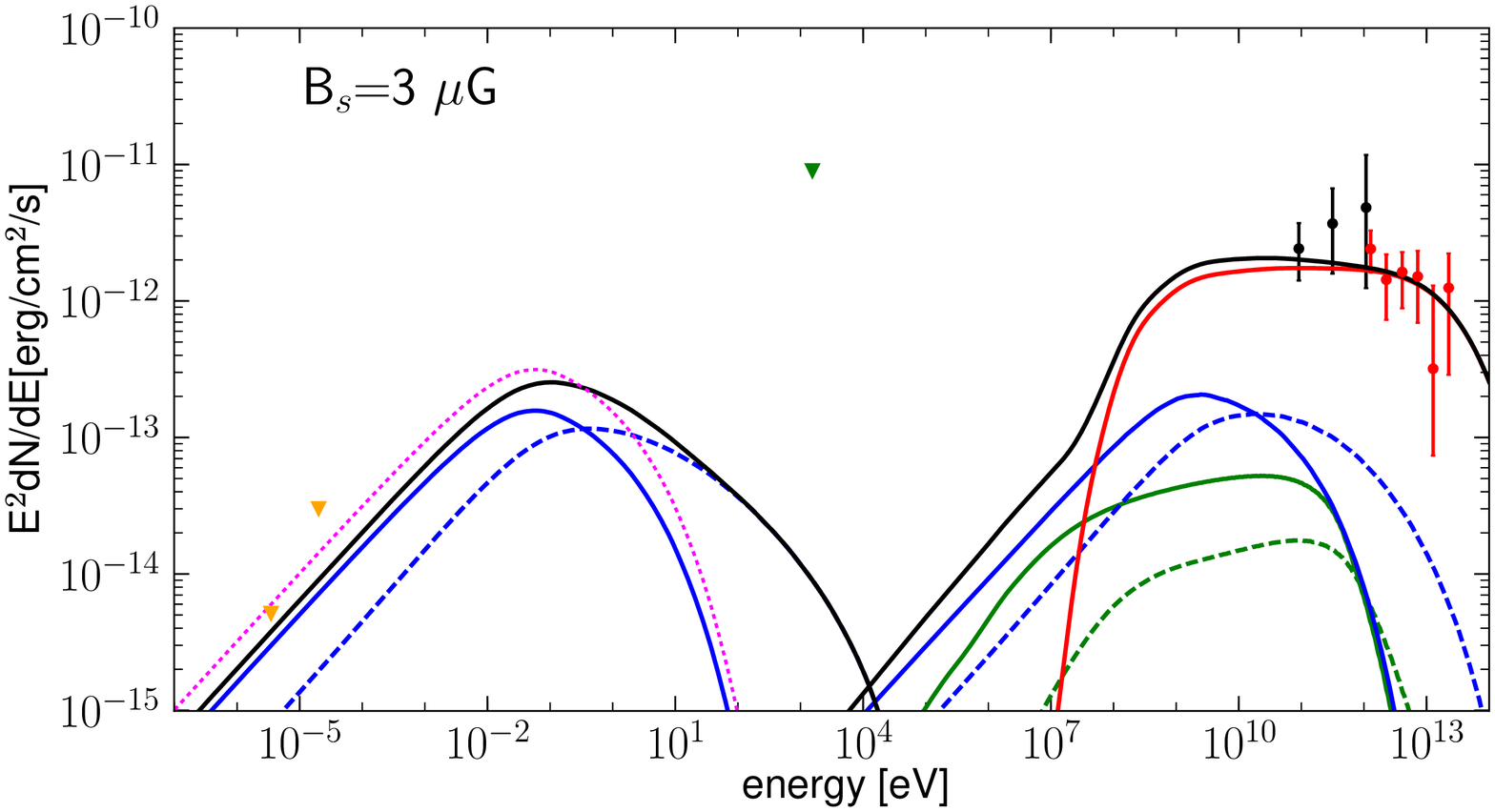}
\caption{
Broadband spectrum of HESS J1503-582 compared with the HVC accretion model.
Observed data are from H.E.S.S. (red) and {\it Fermi}-LAT (black).
Upper limits are from MGPS at 843 MHz and PMN at 4.8 GHz (yellow triangles), and RASS at 1-2.4 keV (green triangle).
Model curves correspond to $\pi^0$ decay emission (red solid),
synchrotron and inverse Compton emission from primary electrons (blue solid) and secondary $e^\pm$ (blue dashed),
bremsstrahlung from primary electrons (green solid) and secondary $e^\pm$ (green dashed),
and total of all components (black).
Model parameters are:
$v_s = 300 {\rm \ km \ s^{-1}}$, $M_{\rm HVC} = 10^5 \ {\rm M_\sun}$, $n_{\rm HVC} = 0.15 \ {\rm cm^{-3}}$,
$D = 20 {\rm \ kpc}$, 
$B_s = 3\mu {\rm G}$, $\eta=1$, $\alpha=2$, and $K_{ep} = 0.005$.
The interstellar radiation field at $R = 15 \ {\rm kpc}$ is adopted.
Also shown is the synchrotron emission from primary electrons only for $K_{ep} =0.01$ (magenta dotted).
The total injected proton energy is $W_p = 4.68 \times 10^{51} \ {\rm erg}$, corresponding to $\xi_p=0.052$.
}
\label{fig:B3muG}
\end{figure}

\begin{figure}[!htb]
\epsscale{1.3}
\plotone{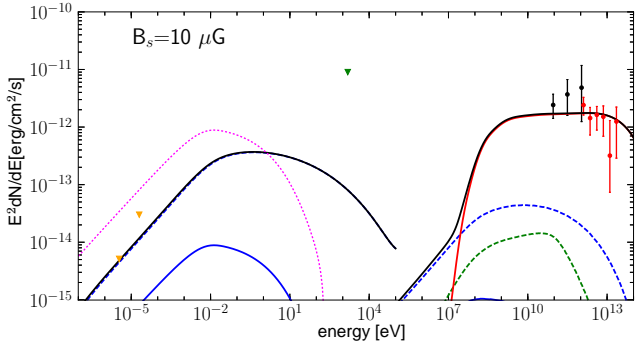}
\caption{
As with Fig. \ref{fig:B3muG},
except for $B_s =10\mu {\rm G}$ and $K_{ep} =10^{-4}$.
The total injected proton energy is $W_p = 4.73 \times 10^{51} \ {\rm erg}$, corresponding to $\xi_p=0.053$.
Also shown is the synchrotron emission from primary electrons only for $K_{ep} =0.01$ (magenta dotted).
}
\label{fig:B10muG}
\end{figure}

\begin{figure}[!htb]
\epsscale{1.3}
\plotone{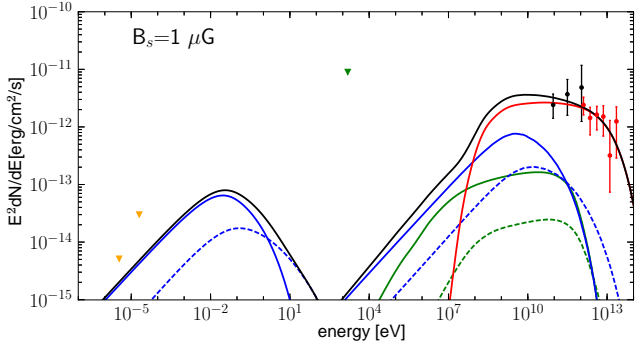}
\caption{
As with Fig. \ref{fig:B3muG},
except for $B_s =1\mu {\rm G}$ and $K_{ep} = 0.01$.
The total injected proton energy is $W_p = 5.1 \times 10^{51} \ {\rm erg}$, corresponding to $\xi_p=0.057$.
}
\label{fig:B1muG}
\end{figure}

\begin{figure}[!htb]
\epsscale{1.3}
\plotone{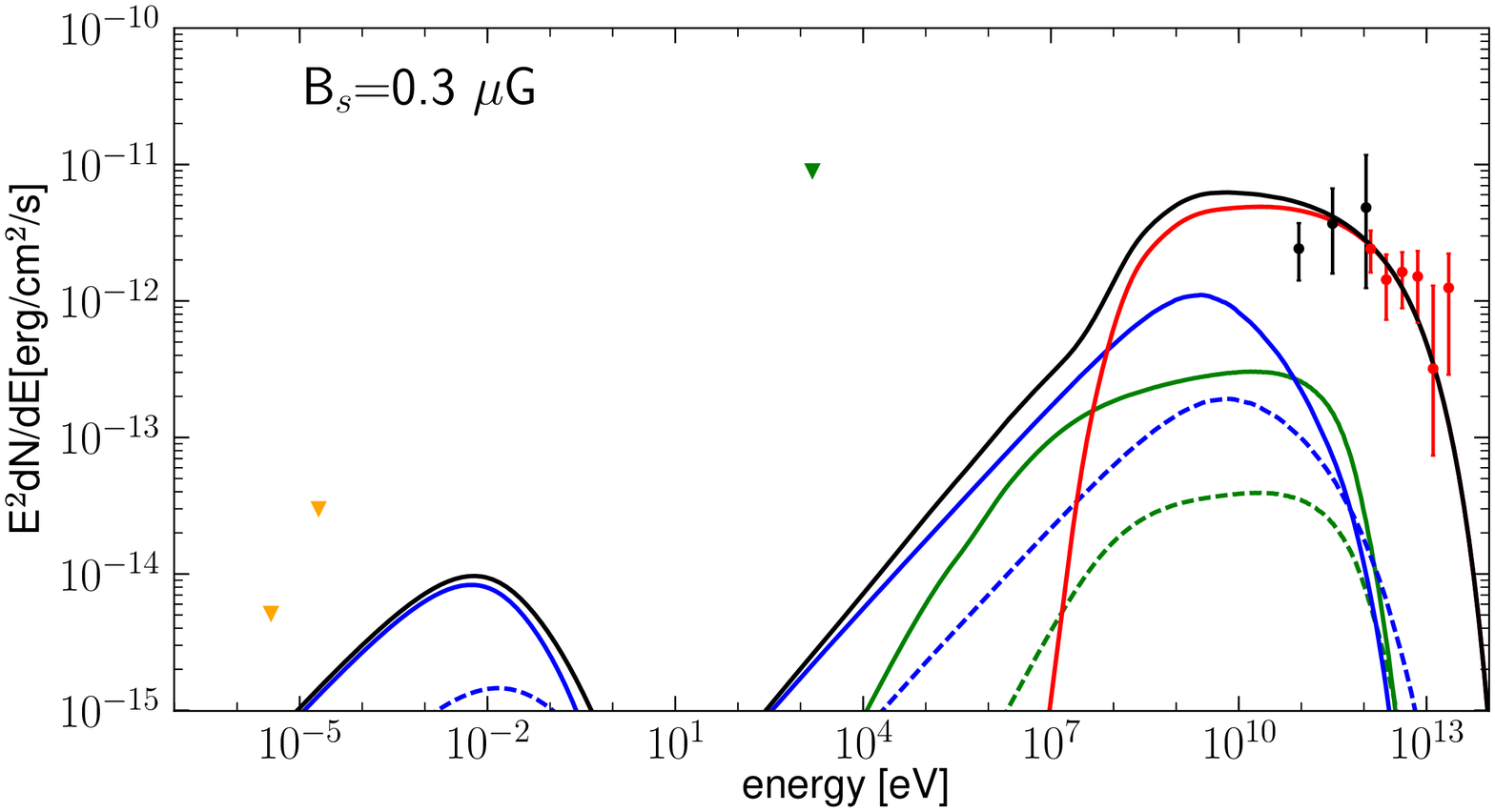}
\caption{
As with Fig. \ref{fig:B3muG},
except for $B_s = 0.3\mu {\rm G}$.
The total injected proton energy is $W_p = 1.1 \times 10^{52} \ {\rm erg}$, corresponding to $\xi_p=0.13$.
}
\label{fig:B03muG}
\end{figure}

Compared with the broadband data of HESS J1503-582
in Figs. \ref{fig:B3muG}, \ref{fig:B10muG}, \ref{fig:B1muG} and \ref{fig:B03muG}
are model results for $B_s =$ 3, 10, 1 and 0.3 ${\rm \mu G}$, respectively.
In order of ascending $B_s$, the corresponding particle maximum energies are
$E_{p,\max} \sim$ 25.6, 85.2, 256, 852 TeV (Eq. \ref{eq:E_qmax}) and
$E_{e,\max} \sim$ 4.7, 8.3, 11.7, 9.7 TeV (Eq. \ref{eq:E_emax}),
which can also be seen in Fig. \ref{fig:energy}.
Given the limited multi-wavelength coverage of the existing data,
all cases with $B_s = 0.3 -10 {\rm \mu G}$ provide generally acceptable descriptions
with plausible values for $\xi_p$ and other parameters.
Nevertheless, these observations already disfavor $B_s < 0.3 {\rm \mu G}$ and $B_s > 10 {\rm \mu G}$, as elaborated below.

The GeV-TeV emission is dominated by the $\pi^0$ component,
whose spectrum agrees fairly well with our estimates in \S \ref{sec:emi_p}.
The inferred cutoff energies are $E_{\pi \gamma,\max} \approx 0.08 \ E_{p,\max} \sim$ 2.1, 6.8, 21, 68 TeV, ordered by $B_s$,
within or somewhat above the energy range covered by H.E.S.S. (Fig. \ref{fig:energy}).
The model is still acceptable for $B_s = 0.3 {\rm \mu G}$ (Fig. \ref{fig:B03muG}),
but $B_s < 0.3 {\rm \mu G}$ is disfavored as the cutoff will become more discrepant with the data at the highest energies.
While the data of \cite{Renaud08} is consistent with an unbroken power-law,
the presence of spectral cutoffs can be tested with further TeV observations (\S \ref{sec:HEobs}).

IC and bremsstrahlung from primary electrons as well as IC from secondary $e^\pm$
can be non-negligible in the GeV band if $B_s \la 3 {\rm \mu G}$.
Due to Klein-Nishina effects, the primary IC spectrum is seen to peak at lower energies compared to our simple estimates in \S \ref{sec:emi_e}.
While the contribution from primary electrons can be made more significant by choosing higher values of $K_{ep}$,
accounting for the observed TeV emission chiefly with such components is problematic,
as $E_{e,\max}$ is limited to $\la$ 12 TeV (\S \ref{sec:acceltime}; Fig. \ref{fig:energy}).

When $B_s \la 2 \ {\rm \mu G}$, $E_{\rm sy,\max} \la 15 \ {\rm eV}$,
reaching only the UV band (\S \ref{sec:emi_e}; Eq. \ref{eq:E_symax}),
unlike SNRs or PWNe for which the synchrotron emission commonly extends to X-ray energies and above.
The radio emission is also faint, below the survey upper limits.
Together with the expectation that the thermal emission from the shock peaks in the UV to soft X-ray bands,
a range most severely affected by photoelectric absorption in the intervening ISM (Eq. \ref{eq:T_s}; \S \ref{sec:thermal}),
this provides one plausible explanation why this source can be ``dark'' (i.e. not readily detectable)
at wavelengths other than gamma rays.

On the other hand, when $B_s \ga 3 {\rm \mu G}$, (Figs. \ref{fig:B3muG}, \ref{fig:B10muG}),
the synchrotron emission can be bright enough to violate the current radio upper limits, unless $K_{ep} \ll 0.01$ is invoked.
We disfavor $B_s \ga 10 {\rm \mu G}$, 
since tension would persist even for the contribution from secondary $e^\pm$ that is independent of $K_{ep}$ (Fig. \ref{fig:B10muG}).
The secondary synchrotron component also extends into and dominates the X-ray band for $B_s \ga 3 {\rm \mu G}$.
While not in conflict with existing limits, 
this may be detectable with deeper observations by current facilities such as {\it Chandra} or {\it XMM-Newton},
which are highly desirable in order to constrain the crucial value of $B_s$ (\S \ref{sec:Xradobs}).

\section{Prospects for further observations}
\label{sec:tests}

We discuss further observations that can
test the HVC accretion origin of dark GeV-TeV sources like HESS J1503-582
and discriminate from more conventional possibilities such as SNRs or PWNe.
Also outlined are strategies to search for further sources triggered by HVC accretion events.

\subsection{HI: morphology, kinematics, distance, location}
\label{sec:HIobs}

As with the elucidation of FVW 40.0+0.5 as a HVC accretion event \citep{Park16},
a pivotal test will be deeper HI observations with higher spatial and spectral resolution
of the region containing FVW 319.8+0.3 and HESS J1503-582 (and other potential candidates),
in order to search for morphological and kinematical evidence of
a HVC, expanding shell, and/or any other sign of HVC-disk interaction.
Significant improvements can be expected in the era of SKA \citep{McClure15}.
We note that the FVW 40.0+0.5 and FVW 319.8+0.3 systems
may be at somewhat different stages of the accretion process.
As discussed by \cite{Park16},
the former may be in a relatively advanced phase, well into the radiative regime,
where most of the kinetic energy initially conveyed by the HVC has been radiated away,
consistent with its estimated age of $5 \times 10^6$ yr (see Eq. \ref{eq:t_grad})
and the small current mass of the HVC (\S \ref{sec:dark}).
In comparison, our interpretation of the latter at a younger age of $t \sim 10^6$ yr
implies that it is only nearing the transition from adiabatic to radiative (\S \ref{sec:shockvt}, \S \ref{sec:neutral}).
Thus the conditions of the parent HVC may be rather closer to its pre-impact state,
although this can also depend on the collision parameters and geometry \citep{Baek08}.

The kinematics information from HI observations
can also provide valuable constraints on the distance.
HVCs may require $M_{\rm HVC} \ga 10^{4.5} M_\sun$
to survive disruption in the halo before reaching the disk (\S \ref{sec:intro}, \S \ref{sec:HVCdisk}),
and the shock requires at least $v_s \sim 120 \ {\rm km \ s^{-1}}$
to avoid suppression of particle acceleration by neutral-ion collisions (\S \ref{sec:neutral}).
Thus, the kinetic energy of HVCs of interest for non-thermal emission is
minimally $E_{\rm HVC,\min} \sim 5 \times 10^{51} \ {\rm erg}$ (Eq. \ref{eq:E_HVC}),
and likely larger, as with our fiducial model parameters (\S \ref{sec:dark}).
If distance constraints point to such values for the required energy,
it can rule out origins related to SNRs or PWNe that are typically less energetic.

Improved constraints on the system's sky position and distance
will give better knowledge of its relative location in the Galaxy,
also offering crucial clues on its origin.
A unique property of HVC accretion events is that their locations with respect to Galactic star forming regions
can be either weakly correlated, totally uncorrelated, or even anti-correlated, depending on the HVC's provenance.
This is in stark contrast to most other types of proposed gamma-ray emitters including SNRs and PWNe
that should be closely connected with star formation.
In particular, HVCs from the IGM and satellites
are initially oblivious to conditions of the disk locations where they accrete,
which may include regions between spiral arms as well as outside the known stellar disk.
In fact, due to their typically higher angular momenta relative to the disk gas,
they are more likely to accrete onto the outer regions of the Galaxy (\S \ref{sec:HVCacc}),
where the star formation rate is considerably reduced.
Furthermore, SN- or SMBH-driven outflows can impede direct HVC accretion more strongly in the inner Galaxy.
These effects may combine to result in a global anti-correlation.
Even on smaller scales, localized outflows from individual star forming regions
may have disruptive effects on HVCs falling close to them, possibly causing local anti-correlations.
If some HVCs originate from Galactic fountains, their launching sites should be star forming regions,
but not necessary their fallback locations, which may also be anti-correlated due to disruption by outflows from the disk.

Note that HVC-disk collisions have actually been proposed to induce star formation \citep[e.g.][]{Franco88},
which may be supported by some observations of the outer Galaxy \citep{Kobayashi08, Izumi14}.
However, this is expected to occur well after the adiabatic phase of our interest,
when sufficient amounts of gas has been swept up and can cool significantly.
Thus the gamma-ray emitting phase of the event is unlikely to be concurrent with the phase of self-induced star formation.

\subsection{GeV-TeV gamma-rays: spectra, localization, morphology}
\label{sec:HEobs}

While not providing definitive proof of its origin,
some expected features of the HVC accretion model can be tested by further GeV-TeV observations.
The high-energy cutoff of the $\pi^0$-dominant gamma-ray spectrum reflects $E_{p,\max}$,
which is determined primarily by $v_s$, $\tau_s$ and $B_s$ (Eq. \ref{eq:E_qmax}).
The magnitude of $v_s$ is controlled by the gravitational potential of the Galaxy
and cannot be much larger than our fiducial value, implying the same for $\tau_s$
that depends mostly on $v_s$ (Eqs. \ref{eq:t_HVC}, \ref{eq:t_disk}, \ref{eq:t_grad}).
Since $B_s < B_{s, {\rm eq}} \sim 14 {\rm \mu G}$ (\S \ref{sec:mag}),
high-energy spectral cutoffs are expected somewhere in the range $\sim 2 - 100$ TeV,
as apparent in Fig. \ref{fig:energy} and Figs. \ref{fig:B3muG} - \ref{fig:B03muG}.
Further observations by current imaging atmospheric Cherenkov telescopes (IACTs)
such as H.E.S.S., MAGIC and VERITAS,
ground-based arrays such as HAWC \citep{Abeysekara17},
and future facilities such as CTA \citep{Acharya13} and LHAASO \citep{DiSciascio16} are essential,
in order to measure the cutoff and constrain key parameters such as $B_s$.

At lower energies, measurement of the ``pion bump'' around $E_{\gamma,\pi} \simeq 70$ MeV
will verify the $\pi^0$ nature of the gamma-ray emission,
an expectation in common with some SNRs \citep{Ackermann13} but not PWNe.
The current {\it Fermi}-LAT data below a few GeV is complicated by a point source as listed in the 3FGL catalog \citep{Acero15}
that may be separate from the 2FHL counterpart of HESS J1503-582,
making difficult a robust characterization of the spectrum at these energies.
Future instruments with improved localization capabilities such as {\it e-ASTROGAM} \citep{DeAngelis16}
will be advantageous.

Better knowledge of the morphology will be important at any energy,
which may be achievable by {\it e-ASTROGAM} below GeV and CTA above few 100 GeV.
At most wavelengths, the spatial distribution of SNRs is generally expected to be shell-like while that of PWNe is center-filled (plerionic).
HVC accretion events may be more complex,
possibly being either shell-like or plerionic, both types simultaneously, or undescribable by either type,
depending on the relative importance of the FS vs RS and the collision geometry \citep[c.f.][]{Baek08}.
Although the distinction may not be straightforward in reality
as SNRs and PWNe can also have non-trivial morphology depending on their environment,
improved characterization of the morphology will be undoubtedly informative.

\subsection{X-ray and radio: spectra, morphology, Faraday rotation}
\label{sec:Xradobs}

Deeper observations in X-rays and radio continuum
are warranted to search for the non-thermal synchrotron and IC emission from primary and secondary electrons,
whose flux depends sensitively on our key variable $B_s$.
Such constraints will be crucial to corroborate
that $B_s$ is sufficient to accelerate protons up to the inferred maximum energies,
independently of the high-energy cutoff in the gamma-ray spectrum (\S \ref{sec:HEobs}).

In our models for HESS J1503-582 in \S \ref{sec:results},
the synchrotron radio flux can exceed the the sensitivity of the past MGPS and PMN surveys
if $B_s \ga 2 {\rm \mu G}$ and $K_{ep}=0.01$, contributed by primary electrons.
The same is true if $B_s \ga 10 {\rm \mu G}$ irrespective of $K_{ep}$, due to the dominance of secondary electrons.
Dedicated radio observations of this object should be worthwhile even if $B_s$ and/or $K_{ep}$ are lower.
Although the {\it ROSAT} limits are not constraining for these models,
the synchrotron X-ray flux for $B_s \ga 3 {\rm \mu G}$,
dominated by secondary electrons and unaffected by uncertainties in $K_{ep}$,
may be detectable with dedicated observations by current facilities such as {\it Chandra} and {\it XMM-Newton}.
Note that a small fraction of the spatial extent of HESS J1503-582 that includes the X-ray source AX J150436-5824
has been observed with {\it Chandra} ACIS-S, resulting in the detection of a faint point-like source \citep{Anderson14}.

If such non-thermal X-ray and/or radio emission from HVC accretion events can be detected with sufficient significance,
the higher angular resolution achievable in these wavebands compared to gamma-rays
can be beneficial for the purpose of localization and morphological characterization,
potentially allowing more detailed studies of spatial correlations, e.g. with HI properties (\S \ref{sec:HIobs})
and distinction from other classes of non-thermal emitters (\S \ref{sec:HEobs}).

Independent constraints on magnetic fields from Faraday rotation of polarized background sources will be very valuable,
for both the HVC and the outer Galactic disk (\S \ref{sec:mag}).
Sufficiently precise measurements may be challenging for regions in the Galactic Plane, 
but considerable progress is anticipated in the future with SKA \citep{Haverkorn15_SKA}.

\subsection{Other wavelengths: thermal signatures}
\label{sec:thermal}

Assuming for simplicity the shocked volume $V_s$ to be a spherical region of radius $r_s$,
the post-shock gas heated to $T_s \simeq 1.3 \times 10^6 \ {\rm K} \simeq 0.11 \ {\rm keV}$ (Eq. \ref{eq:T_s})
will give rise to thermal bremsstrahlung (free-free) emission at photon energy $\epsilon > k_B T_s$ with energy flux
\begin{multline}
  \epsilon f_{\rm ff}(\epsilon) = {32 \sqrt{2} \pi e^6 \over 3 h m_e^{3/2} c^3} \epsilon^{1/2} \exp\left(-{\epsilon \over k_B T_s}\right) (r_c n_0)^2 {V_s \over 4 \pi D^2} \\
  \simeq 4.9 \times 10^{-9} \ {\rm erg \ cm^{-2} s^{-1}} \left(\epsilon \over 0.1 \ {\rm keV}\right)^{1/2} \exp\left(-{\epsilon \over k_B T_s}\right) \\
  \times \left({r_c n_0 \over 0.4 \ {\rm cm^{-3}}} \right)^2 \left(r_s \over 200 \ {\rm pc}\right)^3 \left(D \over 20 \ {\rm kpc}\right)^{-2}
  \label{eq:flux_ff}
\end{multline}
where $e$ is the elementary charge,
the Gaunt factor $g_{ff} = (3 k_B T_s/ \pi \epsilon)^{1/2}$ is adopted as appropriate for $T_s > 13.6 \ {\rm eV}$ and $\epsilon > k_B T_s$,
and only the contribution from fully ionized hydrogen is considered \citep{Rybicki79, Draine11}.

Emission in the UV to soft X-ray bands
is attenuated during propagation in the ISM by a factor $\exp[-\tau_{\rm pe}(\epsilon)]$,
where $\tau_{\rm pe}(\epsilon) = \sigma_{\rm pe}(\epsilon) N_{\rm H}$ is the optical depth to the source,
and $\sigma_{\rm pe}(\epsilon)$ is the effective cross section for photoelectric absorption by all elements in the ISM,
which increases steeply with decreasing $\epsilon$ down to 13.6 eV \citep[e.g.][]{Kaastra08}.
Thus the observable spectrum peaks in a certain range of $\epsilon$, being exponentially suppressed both above and below.
Taking $N_{\rm H} \simeq 1.5 \times 10^{22} \ {\rm cm^{-2}}$
as appropriate for HESS J1503-582 at our assumed distance $D=20$ kpc (\S \ref{sec:dark}),
$\tau_{\rm pe} \simeq$ 10, 3.2 and 0.63
at $\epsilon \sim 0.5$, 1 and 2 keV, respectively.
Combined with Eq. \ref{eq:flux_ff},
we estimate $\epsilon f_{\rm ff}(\epsilon) \exp[-\tau_{\rm pe}(\epsilon)] \ga 10^{-15} \ {\rm erg \ cm^{-2} s^{-1}}$
for $0.5 < \epsilon < 1.7$ keV,
peaking at $\sim 8.1 \times 10^{-14} \ {\rm erg \ cm^{-2} s^{-1}}$ for $\epsilon \sim$ 0.9 keV.
Note that these values are sensitive to even small changes in $T_s$ or $N_{\rm H}$.
Considering the very narrow spectrum, spatial extension of $\sim 0.5 ^\circ$ and source confusion effects in the Galactic Plane,
detecting this with current facilities such as {\it Chandra} or {\it XMM-Newton} could be difficult.
Nevertheless, if similar sources are found in the future at distances or locations with much less $N_{\rm H}$,
for example, in suitable regions in the second or third Galactic quadrants,
detection of the thermal continuum might still be feasible and provide strong support for a HVC accretion event.

Although a detailed discussion is beyond the scope of this paper,
various other types of thermal signatures including emission lines also provide valuable diagnostics.
Compared to many other known classes of interstellar shock phenomena,
such features are generally expected to be weaker here
due to the lower density and lower metallicity of both the HVC and the outer Galactic disk,
especially those related to molecules or dust.
On the other hand, with $Z/Z_\sun \sim 0.2$ and $T_s \sim 10^6 \ {\rm K}$,
metals such as Fe still play a dominant role as gas coolants \citep{Draine11},
and the likelihood that our shocks are moderately radiative (\S \ref{sec:shockvt}, \S \ref{sec:neutral})
entails some level of associated line emission,
in addition to certain transitions of hydrogen and helium.
Subject to interstellar attenuation as described above,
their detection may be challenging but would be extremely valuable,
potentially allowing direct confirmation of the shock velocities
characteristic of HVCs,
not to mention further important constraints on the source distance.

\subsection{High-energy neutrinos}
\label{sec:neuobs}

Detection of high-energy neutrinos from the source would offer unequivocal proof
that protons are accelerated therein to energies approaching PeV,
if not more detailed information regarding its origin.
With our HVC accretion model parameters for HESS J1503-582,
neutrinos are produced via $pp$ collisions up to maximum energies $E_{\nu,\max} \approx 0.04 \ E_{p,\max} \sim$ 1- 34 TeV
for $B_s = 0.3 -10 {\rm \mu G}$
(Fig. \ref{fig:energy}; \S \ref{sec:HEobs}).
For $B_s > 3 {\rm \mu G}$ and $D= 20 \ {\rm kpc}$,
the estimated neutrino flux per flavor at $E_\nu \sim 10 \ {\rm TeV}$
is $E_\nu^2 d{\cal N}_{\nu,pp} / dE_\nu \simeq 1.8 \times 10^{-12} \ {\rm erg \ cm^{-2} s^{-1}}$ (Eq. \ref{eq:flux_nu}).
At $E_\nu \la 40$ TeV,
the angular resolution of IceCube for through-going track events is $\ga 0.5 ^\circ$,
so sources similar to HESS J1503-582 can be considered point-like.
Compared with the 5$\sigma$ discovery potential of IceCube for point sources of muon neutrinos at $E_\nu \ga 10 \ {\rm TeV}$
\citep{IceCube17_source},
the predicted flux is about an order of magnitude below for a northern source at declination $\delta > 0 ^\circ$,
and up to 2-3 orders of magnitude below for a southern source such as HESS J1503-582.
Nevertheless, if similar sources can be found in the northern hemisphere in the future, they could still be interesting for IceCube,
especially if its sensitivity at lower energies can be enhanced \citep{IceCube_PINGU}.
Better yet, realistic prospects of detecting neutrinos from southern Galactic sources including HESS J1503-582
can be foreseen for KM3NeT \citep{KM3NeT}.

\subsection{Searches for HVC accretion events}
\label{sec:search}

As discussed in \S \ref{sec:enenum},
the expected total number of HVC accretion events in the Galaxy
with shocks in the adiabatic phase producing multi-TeV emission is fiducially $N_s \sim 10$,
but can be larger or smaller depending on
the unknown distribution of $v_{\rm acc}$ and $M_{\rm HVC}$.
If such shocks continue to accelerate particles into their radiative phase,
a larger number of sources with sub-TeV emission may result,
perhaps with a spectral break, as discussed in some models for old SNRs (\S \ref{sec:neutral}).
Searches are warranted for further candidate HVC accretion events
in all wavebands and channels discussed above, especially at GeV-TeV energies and in HI.

Unidentified sources in the Galactic Plane found by H.E.S.S. and other IACTs, HAWC, and {\it Fermi}-LAT
should be scrutinized for correlations with prominent HI structures such as FVWs or large shells,
which will be facilitated in the future by CTA and {\it e-ASTROGAM} with their higher angular resolution.
Similar studies for high-energy neutrinos detected near the Galactic Plane may also be interesting.
Alternatively, one can select particularly interesting FVWs or other objects with noteworthy HI properties 
and perform targeted IACT observations,
keeping in mind the expectations noted in \S \ref{sec:HIobs}
such as likely location in the outer Galaxy with weak to opposite correlation with star forming regions.
We note that while the CHVC+supershell system of \cite{Park16} may be deep into its radiative phase
and possibly less efficient as a particle accelerator (\S \ref{sec:HIobs}),
dedicated observations of the region may still be worthwhile.
The available HI data should improve dramatically with the advent of SKA \citep{McClure15}.

A very intriguing possibility is the detection of HVC accretion events in external galaxies such as M31.
HVCs have been observed within projected distance $\sim$ 100 kpc of M31
with estimated HI masses $M_{\rm HI} \sim 10^5 - 10^7 M_\sun$ \citep{Thilker04, Westmeier08, Lockman17}.
If a relatively large HVC of total mass $M_{\rm HVC} \sim 5 \times 10^6 M_\sun$
is accreting at $v_{\rm acc} \sim 300 \ {\rm km \ s^{-1}}$ onto the the disk of M31 at distance $D \simeq 780 \ {\rm kpc}$,
protons shock-accelerated therein to $E_p \sim 50 \ {\rm TeV}$
can give rise to $pp$ $\pi^0$ gamma-ray emission at $E_\gamma \sim 4 \ {\rm TeV}$
with energy flux $E_\gamma^2 d{\cal N}_{\gamma,pp} / dE_\gamma \simeq 1.1 \times 10^{-13} \ {\rm erg \ cm^{-2} s^{-1}}$
and spatial extension $\theta \sim 2 r_{\rm HVC} / D \simeq 0.12 ^\circ$,
taking the same parameters as in Eqs. \ref{eq:r_HVC} and \ref{eq:flux_pp} except for $M_{\rm HVC}$, $L_{k, {\rm HVC}}$, $E_p$, and $D$.
This may be within reach of the sensitivity and angular resolution of CTA North \footnote{\url{https://www.cta-observatory.org}},
potentially offering unique information on HVC accretion in a disk galaxy other than our own,
even though the likelihood of such an energetic event occurring at a given time may not be large
($\sim$ 20 \% from Eq.\ref{eq:N_s} if $\dot{M}_{\rm acc, HVC}$ for M31 is similar to the Milky Way).
Note that {\it Fermi}-LAT has detected an extended source around the center of M31
with radius $\sim 0.4 ^\circ$,
energy flux $\sim 5.6 \times 10^{-12} \ {\rm erg \ cm^{-2} s^{-1}}$ at 0.1-100 GeV,
and a power-law spectrum with photon index $\sim$2.4, which may or may not be diffuse interstellar emission induced by CRs \citep{Ackermann17_M31}.
Even if the spectrum of this {\it Fermi} source extends unbroken into the TeV band, 
confusion with the HVC accretion event considered above should not be an issue
if the latter's location is farther than 5.5 kpc projected distance from the center of M31.

\section{Conclusions and prospects}
\label{sec:conc}

As observationally established facts:\\
1. The Galactic disk is accreting low-metallicity gas at a total rate of order $\sim 1 \ M_\sun {\rm yr^{-1}}$ (\S \ref{sec:gasacc}).\\
2. At least part of this accretion proceeds in the form of high velocity clouds of cool gas with 
mass $\sim 10^5 M_\sun$ and mean gas density $\sim 0.1 \ {\rm cm^{-3}}$
directly impacting the outer regions of the disk at velocities of a few 100 ${\rm km \ s^{-1}}$,
as seen in an object initially identified as a forbidden velocity wing (\S \ref{sec:HVCacc}, \S \ref{sec:dark}).\\
3. Numerous GeV-TeV sources in the Galactic Plane are spatially extended and unidentified,
of which at least one, HESS J1503-582, is dark (i.e. undetected in any other waveband)
except for spatial association with a forbidden velocity wing of unknown physical nature (\S \ref{sec:HVCnt}, \S \ref{sec:unID}, \S \ref{sec:dark}).\\

Facts 1 and 2 entails the formation of collisionless adiabatic shocks with lifetime $\sim 10^6 \ {\rm yr}$ (\S \ref{sec:shock}).
Assuming shock velocity $v_s \sim 300 \ {\rm km \ s^{-1}}$ and magnetic fields of order a few $\mu$G,
protons and electrons can be accelerated up to sub-PeV and multi-TeV energies, respectively (\S \ref{sec:accel}),
resulting in GeV-TeV gamma-ray emission primarily via p-p $\pi^0$ decay with some additional contribution from inverse Compton
(\S \ref{sec:emission}).
Part of Fact 3 concerning HESS J1503-582 can be consistently and plausibly accounted for in such terms.
Despite the currently limited multi-wavelength data, 
the observed TeV spectrum and upper limits on radio synchrotron emission from secondary $e^\pm$
imply the constraint $B_s \sim 0.3 - 10 \ {\rm \mu G}$ (\S \ref{sec:results}).

Further observational tests of HVC accretion events as non-thermal emitters include:
better HI observations of suitable GeV-TeV sources to clarify their morphology and kinematics,
and to constrain their distances and locations that could have little, no or opposite correlation with star-forming regions (\S \ref{sec:HIobs});
deeper X-ray and radio observations of such sources aiming for synchrotron components and constraints on $B_s$ (\S \ref{sec:Xradobs});
better GeV-TeV observations (\S \ref{sec:HEobs});
and searches for thermal signatures (\S \ref{sec:thermal}) and high-energy neutrinos (\S \ref{sec:neuobs}).
Systematic studies are worthwhile to hunt for more such sources in surveys of the Galactic Plane at GeV-TeV and in HI,
and possibly even in external galaxies like M31 (\S \ref{sec:search}).

Our theoretical formulation can be improved in various respects.
For shock properties, the lack of symmetry in the problem calls for
3-D hydrodynamical simulations including radiative cooling effects
for a better description of its dynamical evolution and dependence on collision parameters (\S \ref{sec:shockvt}).
The plasma physics of particle acceleration at the low shock velocities of our interest, as well as in the radiative phase not treated here,
is worth exploring through particle-in-cell simulations,
including the potential effects of magnetic field amplification (\S \ref{sec:mag})
and charge exchange reactions induced by neutral particles (\S \ref{sec:neutral}).

We have focused on non-thermal emission induced by CRs at the source,
and have not explicitly addressed the consequences of CRs escaping from them.
While the contribution of HVC accretion to the total Galactic CR budget is estimated to be minor,
maximally 15 \% (\S \ref{sec:enenum}),
it can still be relatively important in the outer disk regions,
where the occurrence of conventional CR sources like SNRs are much rarer
(\S \ref{sec:HVCacc}, \S \ref{sec:HIobs}).
For example, compared to $R \sim 4 \ {\rm kpc}$ where Galactic star formation peaks,
the star formation rate as traced by SNRs or pulsars is $\ga 10$ times lower at $R \ga 15 \ {\rm kpc}$ \citep{Stahler05}.
Considering the likelihood that the radial dependence of gas accretion is biased outwards
relative to star formation in the disk \citep{Peek09, Christensen16, Stewart17},
CRs of HVC accretion origin may even be dominant in the outer Galaxy.
This can have interesting implications for the diffuse Galactic gamma-ray emission,
whose observed intensity at $R>R_\sun$ has long known to be in excess
of expectations based on CR sources that follow star formation \citep{Abdo10, Ackermann11,Ackermann12, Acero16}.
The observed contrast between spiral arm and inter-arm regions also appears to be weaker than expected
for CR production tracing star formation \citep{Ackermann11, Grenier15}.
Proposed solutions to the discrepancy include non-conventional modes of CR diffusion \citep[e.g.][]{Evoli12},
and dark gas missed by existing observations \citep{Grenier15}.
The intriguing prospect that CRs induced by HVC accretion is behind this mystery
will be discussed in a future publication.

Finally, we touch on the possibility that future observations of non-thermal phenomena triggered by accretion of cool gas
may provide a fresh perspective on studying gas accretion onto galaxies per se,
many aspects of which are still poorly understood \citep{Fox17}.
As discussed above,
HVC accretion events are objects that may not be readily detectable at wavelengths other than GeV-TeV gamma rays,
with heavily attenuated thermal emission (\S \ref{sec:thermal}),
possibly accompanied by weak and diffuse radio and X-ray emission (\S \ref{sec:results}, \S \ref{sec:Xradobs}).
Their clear identification may only be feasible by HI observations with high angular and spectral resolution (\S \ref{sec:HIobs}).
While GeV-TeV emission does not provide unique indication of HVC accretion
nor detailed information on properties of the parent HVC,
they do potentially serve as signposts that illuminate the accretion interface
and are visible across the Galaxy to its outer edges,
and perhaps also in the nearest external galaxies such as M31 (\S \ref{sec:search}).
If at least some GeV-TeV sources can be revealed to be HVC accretion events
via follow-up observations in HI and other wavelengths,
the same observations should constrain their distance and location,
and thereby provide information on key parameters of the HVC such as its mass, velocity, density and temperature.
Achieving a sufficient sample of such observations at different locations in the Galaxy
could allow characterization of the distribution of these parameters and its spatial dependence,
which is highly uncertain at present (\S \ref{sec:HVCdisk}).
This in turn could offer potential discrimination of the source(s) of the accreting gas,
among IGM filaments, satellite galaxies, halo condensation, etc.
In addition, non-thermal emission offers direct constraints on magnetic fields,
which may not be easily obtainable otherwise.
As accretion of cold gas is likely more efficient at early epochs (\S \ref{sec:gasacc}),
related effects that may have occurred in the past and left observable traces,
e.g. CR-induced production of light elements \citep{Suzuki02}, are also worth exploring.
Such studies can offer a novel approach to probe
gas accretion processes and the evolutionary cycle of baryons in the Milky Way and other galaxies.

\acknowledgments
We are grateful to Changhyun Baek for his contributions during the early stages of this work.
Valuable discussions are acknowledged with
Takahiro Kudoh, Yutaka Ohira, Ryo Yamazaki, Masahiro Nagashima and Masaki Mori.
This work is supported by JSPS KAKENHI Grant Numbers JP17K05460 (SI), JP26247027 (YU), and 16H03959 (KW),
and by the RIKEN Junior Research Associate Program (MA).

\bibliography{Gacc-CR_rev}

\end{document}